\documentclass[11pt,final,onecolumn]{IEEEtran}
\usepackage[latin1]{inputenc}
\usepackage[francais]{babel}
\usepackage{amsmath}
\usepackage{amssymb,amsmath}
\usepackage{setspace}
\usepackage{graphicx}
 
\newcommand{\Argmin}[1]{\mbox{Arg}\min_{#1}}

\newcommand{\hh}{\mathcal{H}}

\newcommand{\pds}[2]{\Big|\Big<#1, #2\Big>\Big|}
\newcommand{\pdp}[2]{\Big<#1, #2\Big>}

\begin{document}

\title{Compressed Sensing in Astronomy}
\author{J. Bobin${}^{\star}$, J-L.Starck and R. Ottensamer 
\thanks{J.Bobin${}^{\star}$  (E-mail: jerome.bobin@cea.fr) and J-L.Starck (E-mail: jstarck@cea.fr) are with the
Laboratoire AIM, CEA/DSM-CNRS-Universit\'e Paris Diderot, CEA Saclay, IRFU/SEDI-SAP,
Service d'Astrophysique, Orme des Merisiers, 91191 Gif-sur-Yvette, France. Phone:+33(0)169083118. Fax:+33(0)169086577.}
\thanks{R.Ottensamer  (E-mail: ottensamer@astro.univie.ac.at) is with the University of Vienna, Institute of Astronomy, T\"urkenschanzstr. 17, A-1180 Wien, Austria.}
\thanks{This work is partly supported by the Austrian Federal Ministry of Transport, Innovation and Technology within the project FIRST/PACS Phase I and the ASAP project of the FFG/ALR.}}

\maketitle

\def\ff{FITSPRESS}
\def\jj{JPEG}
\def\hh{HCOMPRESS}
\def\co{compression}
\def\pmt{PMT}
\def\MM{MathMorph}
\def\sext{SExtractor}
\def\MVM{Multiscale Vision Model}

\begin{abstract}
Recent advances in signal processing have focused on the use of sparse representations in various applications. 
A new field of interest based on sparsity has recently emerged~: \textit{compressed sensing}. 
This theory is a new sampling framework that provides an alternative to the well-known Shannon sampling theory. In this paper we investigate how \textit{compressed sensing} (CS)  can provide new insights into astronomical data compression and more generally how it paves the way for new conceptions in astronomical remote sensing.
We first give a brief overview of the compressed sensing theory which provides very simple coding process with low computational cost, thus favoring its use for real-time applications often found on board space mission. We introduce a practical and effective recovery algorithm for decoding compressed data.  In astronomy, physical prior information is often crucial for devising effective signal processing methods. We particularly point out that a CS-based compression scheme is flexible enough to account for such information. In this context, \textit{compressed sensing} is a new framework in which data acquisition and data processing are merged. 
We show also that CS provides a new fantastic way to handle multiple observations of the same field view, allowing us to recover information
at very low signal-to-noise ratio, which is impossible with standard compression methods. This CS data fusion concept could
lead to an elegant and effective way to solve the problem ESA is faced with, for the transmission to the earth of the data collected by
PACS, one of the instruments on board the Herschel spacecraft which will launched in 2008.
 
\end{abstract}



\section*{Introduction}
 From year to year, the quantity of astronomical data increases  	
at an ever growing rate.  In part this is due to very large digital sky
surveys in the optical and near infrared, which in turn has been made possible by 
the development of digital imaging arrays such as CCDs (charge-coupled
devices).  The size of  digital arrays is continually growing, pushed by the demands of astronomical 
research for ever larger quantities of data in ever shorter time  	
periods. As a result, the astronomical community is also confronted with  	
a rather desperate need for data compression techniques.    	
Several techniques have in fact been used, or even developed, 	
for astronomical data compression. V\'eran \cite{compress:Veran} studied 
lossless techniques. White et al.\ \cite{compress:white92} developed \hh, 
based on the Haar wavelet transform, and
Press et al.\ \cite{compress:press92} developed  \ff \ based on the 
Daubechies wavelet transform.	In addition, the scientist must of course consider  \jj, a 
general purpose standard. Effective and efficient compression based on the multiresolution 	
 Pyramidal Median Transform (PMT) algorithm was developed by 
Starck et al.\ \cite{starck:sta96_2}.
 Huang and Bijaoui \cite{compress:huang91_1} used mathematical morphology  in  \MM \ for astronomical	
 image processing.

For some projects, we need to achieve 
huge compression ratios, which cannot be obtained by current  methods without 
introducing unacceptable distortions. For instance, it was shown \cite{compress:dollet04}
that if we wish to extend the GAIA mission in order to make a
high-spatial resolution all-sky survey in the visible based on a 
scanning satellite, then the main limitation is the amount of collected data to be transmitted.
A solution could be to introduce all  our knowledge of both the sky and the instrument in order to compress only
the difference between what we know and what we observe  \cite{compress:dollet04}. 
However, errors on the point spread functions, positions of stars, etc., must be under control \cite{compress:dollet04} and the computation
cost on board of the satellite may be unacceptable.  The Herschel satellite\footnote{See \textit{http://www.esa.int/science/herschel}}, which will be launched in 2008, is faced with a similar problem. Indeed the photometer data
need to be compressed by a factor of $16$ to be transferred. The yet implemented lossless compression scheme (based on entropy coding) yield a compression rate of $2.5$. ESA\footnote{See \textit{http://www.esa.int}.} is in need of a compression ratio of $6$. As the CPU load has to be extremely small, conventional compression methods cannot be used. 

Recently, an alternative sampling theory has emerged which shows that signals can be recovered from far fewer samples (measurements) than what the Nyquist/Shannon sampling theory states. 
This new theory coined \textit{compressed sensing} or (compressive sensing)  (CS) introduced in the seminal papers \cite{CRT:cs,CT:cs3,Donoho:cs} relies on the compressibility of signals or more precisely 
on the property for some signals to be sparsely represented. 
In a more general setting, sparsity is known to entail effective estimation (restoration, blind source separation $\cdots$ etc.), efficient compression 
or dimension reduction. From the compressed sensing viewpoint, sparse signals could be acquired ``economically" (from a few samples) \textit{without loss of information}. 
 It introduces new conceptions in data acquisition and sampling. It has been shown that CS could be useful in many domains such as medical imaging \cite{Lustig07}, biosensing \cite{Sheikh07}, radar imaging \cite{Baradar07}
 or geophysical data analysis \cite{LHerm07}. \\
 
 \textit{Scope of the paper :} We propose a new alternative approach for the transmission of astronomical images, based on CS. 
 Similarly to classical compression schemes, CS can be arranged as a ``Coding-Decoding" two-stage scheme. In practical situations (more particularly for on board applications), 
 CS provides a particularly simple coding stage that only requires a low computational cost. Most of the computational complexity is then carried by the decoding step. 
 In this context, we introduce a new decoding algorithm that quickly and accurately provides close solutions to the decoding problem. 
 Section~\ref{sec:sect_sensing} reviews the principle of the CS theory.  Section~\ref{sec:cs_astro} shows how CS can be used in astronomy and presents a decoding algorithm.
More generally, we introduce a new conception of astronomical remote sensing; we particularly show that the CS framework is able to account for specific physical priors. 
 It paves the way for new instrument design in which data acquisition, compression and processing can be merged. 
In section~\ref{sec:csomp} we show how CS offers us a new data fusion framework when multiple observations of the same field of view are 
 available. This happens very often in astronomical imaging when we need to build a large map from a  micro-scan or a raster-scan strategy.
 Section~\ref{sec:results} emphasizes on the effectiveness of the proposed  CS-based compression for solving the Herschel data compression problem. Indeed, we show the advantage of CS over the averaging approach which has been considered so far.



\section{An overview of compressed sensing theory}
\label{sec:sect_sensing}

In this section, we give a brief and non exhaustive review of compressed sensing and show how this new sampling theory will probably lead to a ``revolution" in signal processing and communication theory. For more exhaustive tutorials in this field, we refer the reader to the review papers \cite{Candes:cs1,Candes:2007lq}.
Assume $x \in \mathbb{R}^t$ (written as a column vector with $t$ entries) such that we ``observe" or ``measure" only $M < t$ samples $\{y_k\}_{k=1,\cdots,M}$. These measures are obtained by projecting the signal $x$ on  a set of so-called measurement vectors $\{\theta_k\}_{k=1,\cdots,M}$ as follows~:
\begin{equation}
y_k = \pdp{x}{\theta_k}
\end{equation}
The backbone of compressed sensing relies on two major concepts~: i) the data to compress are indeed compressible; more precisely the data $x$ have a ``structured" content  so that they can be sparsely represented in some basis $\bf \Phi$; ii) the measurement vectors $\{\theta_k\}_{k=1,\cdots,M}$ are non adaptive (they should not depend on $x$) and incoherent with the basis in which $x$ is assumed to be sparse.

\subsection{The gist of compressed sensing}
\label{sec:csgist}

\paragraph*{Compressibility}

Most ``natural" images or signals have highly structured contents (\textit{i.e.} contours and textures in image processing). Recent advances in harmonic analysis have provided tools that efficiently represent such structures (wavelets, ridgelets \cite{cur:candes99_1}, curvelets \cite{curv:demanet,starck:sta01_3}, contourlets \cite{cur:do05}, to name a few). In this context, efficient representations mean sparse representations. Let's consider a signal $x$ of size $t$. Assume that $x$ can be represented from $T \geq t$ signal waveforms $\{\phi_i\}_{i=1,\cdots,T}$~:
\begin{equation}
x = \sum_{i=1}^T \alpha_i \phi_i
\end{equation}
This relation can be more conveniently recast in matrix formulation~: $x = {\bf \Phi} \alpha$. 
The signal $x$ is said to be sparse in $\bf \Phi$ if most entries of the so-called coefficient vector $\alpha$ are zero or close to zero and thus only a few have significant amplitudes. Such signal $x$ can be efficiently approximated (with low $\ell_2$ approximation error) from only a few significant coefficients. In the extreme case, $x$ is $K$-sparse~: $x$ can be exactly synthesized from $K \ll t$ coefficients. Then such sparse signal is highly \textit{compressible} as the knowledge of only $K$ parameters is needed to perfectly reconstruct the signal $x$.\\ 
Note that, in the last decade, sparsity has emerged as one of the leading concepts in a wide range of signal processing applications (restoration \cite{StarckCD02}, feature extraction \cite{starck:sta05_4}, source separation \cite{bobin:gmca_itip}, compression (\cite{cur:vetterli01,Taubman:2001ij}), to name only a few).\\
Recently, a wide range of theoretical and practical studies have focused on sparse decomposition problems in overcomplete (the case $T > t$) signal waveform dictionaries (see \cite{Donoho-Bruckstein-Elad} and references therein). In this paper we will mainly focus on sparsity assumptions in orthonormal bases $\bf \Phi$. Extensions to overcomplete dictionary would be straightforward in the light of the aforementioned references.\\
From now we assume that $x$ have a $K$-sparse decomposition in the orthobasis $\bf \Phi$. The data $x$ are then compressible; the next problem then amounts to accounting for signal compressibility to devise efficient \textit{non-adaptive} signal compression.

\paragraph*{Incoherence of the measurements}

As an intensive field of research, several works have already addressed compressed sensing in various settings (see \cite{DT:cs,CRT:cs,CR:cs3} and references therein). In the aforementioned references, the way the measurements are designed plays a crucial role. Let us assume that the signal $x \in \mathbb{R}^t$ is a highly compressible $K$-sparse signal in the orthobasis $\bf \Phi$. In compressed sensing, measurements are simple linear projections $\{y_k\}_{k=1,\cdots,M}$~: $y_k = \pdp{x}{\theta_k}$.  Historical works considered measurements from random ensembles (see \cite{DT:cs,CRT:cs,CT:cs3,CR:cs3} and references therein). In these seminal papers, randomness is likely to provide \textit{incoherent} projections. Recall that the coherence between two matrices is measured by their mutual coherence (see \cite{miki:DonohoHuo,Candes:2007lq}) ~:
\begin{equation}
\mu_{\bf \Theta,\bf \Phi} =  \max_{i,j} \pds{\theta_i}{\phi_j}
\end{equation}
In practical situations, measurement vectors are designed by selecting at random a set (indexed by $\Lambda$) of vectors from a deterministic ensemble $\bf \Theta$ as suggested in \cite{CR:cs4,Candes:2007lq}~: $y = {\bf \Theta}_{\Lambda} x$.


\subsection{Signal recovery}
\label{sec:stcs}

\paragraph{Exact solutions}
The previous paragraph emphasized on the way the coding/sensing step should be devised. The decoding step amounts to recover the original signal $x$ out of the compressed signal $y = {\bf \Theta}_{\Lambda} x$.
Furthermore, $x$ is known \textit{a priori} to be $K$-sparse in $\bf \Phi$~: $x = {\bf \Phi} \alpha$ where $\alpha$ is a sparse vector of size $t$. Then the recovery problem boils down to the following sparse decomposition issue in the overcomplete system ${\bf \Theta}_{\Lambda} {\bf \Phi}$~:
\begin{equation}
\label{eq:l0_rec}
\min_{\alpha} \|\alpha\|_{\ell_0} \mbox{ s.t. } y= {\bf \Theta}_{\Lambda} {\bf \Phi} \alpha
\end{equation} 
In the last decade, sparse decomposition issues have been a very active field. Strong recovery results have been provided (see \cite{miki:DonohoHuo,Donoho-Bruckstein-Elad,GIG}). Classically, the $\ell_0$ norm is substituted with the convex $\ell_1$-norm to avoid the combinatorial nature of the problem in Equation~\eqref{eq:l0_rec}. The recovery problem is then recast in a convex optimization program~:
\begin{equation}
\label{eq:l1_rec}
\min_{\alpha} \|\alpha\|_{\ell_1} \mbox{ s.t. } y= {\bf \Theta}_{\Lambda} {\bf \Phi} \alpha
\end{equation} 
Equivalence between these problems has led to a considerable literature (see \cite{Donoho-Bruckstein-Elad} and references therein). At first sight, the decoding step in compressed sensing is equivalent to a sparse decomposition problem in an overcomplete system $\bf \Psi$. Formally, the specificity of CS relies on the particular structure of the overcomplete representation at hand~: ${\bf \Psi} = {\bf \Theta}_{\Lambda} {\bf \Phi}$. Several strong recovery  results in the particular CS framework have been proved based on specific assumptions with random measurement ensembles  (see \cite{don:cs,DT:cs,CRT:cs, CR:cs2}). In practice, as stated earlier, measurements are more conveniently devised from random subsets of deterministic ensembles. 


\paragraph{Approximate solutions}
In practice, signals are seldom $K$-sparse. Furthermore, the data are often corrupted by noise. A more realistic compression model would be the following~:
\begin{equation}
y = {\bf \Theta}_{\Lambda} (x + n)
\end{equation}
where $n$ is a white Gaussian noise with variance $\sigma_{n}^2$. As the measurement matrix ${\bf \Theta}_{\Lambda}$ is a sub-matrix of the orthonormal matrix ${\bf \Theta}$, the projected noise $n_{\Lambda} = {\bf \Theta}_{\Lambda} n$ is still white and Gaussian with the same variance $\sigma_{n}^2$. The projected data are then recast as follows~: $y = {\bf \Theta}_{\Lambda} x + n_{\Lambda}$.
 The recovery step then boils down to solving the next optimization problem~:
\begin{equation}
\label{eq:l1_app}
\min_{\alpha} \|\alpha\|_{\ell_1} \mbox{ s.t. } \left\|y - {\bf \Theta}_{\Lambda} {\bf \Phi} \alpha \right\|_{\ell_2} \leq \epsilon
\end{equation}  
where $\epsilon$ is an upper bound of $\|n\|_{\ell_2}$. Defining $\epsilon = \sqrt{t + 2\sqrt{2t}}\sigma_n$ provides a reasonable upper bound on the noise $\ell_2$ norm, with overwhelming probability.
This problem is known as the LASSO in statistics \cite{lasso:tib} or Basis Pursuit denoising \cite{wave:donoho98}. In the noiseless case ($\epsilon =  0$), it has been shown in \cite{CRT:cs2} that the solution to the problem in Equation~\eqref{eq:l1_app} leads to an approximation error close to the optimal sparse approximation. The optimal sparse approximations would be obtained by reconstructing $x$ from its $K$ most significant coefficients in $\bf \Phi$ (if they were known !). In the noiseless case, the solution to the problem in Equation~\eqref{eq:l1_app} is also shown to provide stable solutions.\\
The convex program (second-order cone program) in Equation~\eqref{eq:l1_app} then provides an efficient and robust mechanism to provide an approximate to the signal $x$. A wide range of optimization techniques (see \cite{donoho:stgomp,l1magic,ist:fnw} to quote a few) providing fast algorithms have been devised to solve the problem in Equation~\eqref{eq:l1_app}.


\section{compressed sensing in Astronomy}
\label{sec:cs_astro}

In the next sections, we focus on applying the compressed sensing framework to astronomical remote sensing. In Section~\ref{sec:astro_coding}, we show that compressed sensing and more precisely its way of coding information provides alternatives to astronomical instrument design. Section~\ref{sec:astro_decoding} gives emphasis on the ability of CS decoding to easily account for physical priors thus improving the whole compression performances.

\subsection{A new way of coding signals}
\label{sec:astro_coding}
In the compressed sensing framework, the coding step needs a very low computational cost. Compressed sensing is then very attractive in several situations~: i) narrow transmission band (for remote sensing) or/and ii) compressing large amount of data; for instance in fast scanning or wide field sensing.  Indeed, in the compressed sensing framework, the way of coding information can impacts on instrumentation in two ways as detailed hereafter.

\subsubsection{\underline{Measuring physics : }}
The philosophy of compressed sensing (\textit{i.e.} projecting onto incoherent measurement ensembles) should be directly applied on the design of the detector. Devising an optical system that directly ``measures" incoherent projections of the input image would provide a compression system that encodes in the analog domain. \underline{Compression would be made by the sensor itself !}\\
Interestingly, such kind of measurement paradigm is far from being science-fiction. Indeed, in the field of $\gamma$-ray imaging, the so-called coded-masks\footnote{We invite the interested readers to visit the following site that is devoted to coded aperture imaging~: \textit{http://astrophysics.gsfc.nasa.gov/cai/}.} (see \cite{Skinner:2002qd} and references therein) are used since the sixties and are currently operating in the ESA/Integral space mission\footnote{See \textit{http://~sci.esa.int/science-e/www/area/index.cfm~?~fareaid=21}.}. In $\gamma$-ray (high energy) imaging, coded masks are used as aperture masks scattering the incoming $\gamma$ photons. More formally, the couple (coded aperture mask and detector field) is equivalent to selecting some projections in the Fourier space. In coded aperture imaging, the way the mask is designed is likely to simulate incoherent projections. Furthermore, $\gamma$-ray data are often made of point sources that are almost sparse in the pixel domain. Fourier measurements then provide near optimal incoherent projections. The first application of compressed sensing then dates back to the sixties ! In the compressed sensing community, the coded mask concept has inspired the design of the celebrated ``compressed sensing camera" \cite{TLW:cs} that provide effective image compression with a single pixel.\\
In coded aperture imaging, the decoding step is often performed by iterative techniques based on maximum entropy \cite{Strong03}. Applying a sparsity-based recovery technique as advocated by the compressed sensing theory would probably provide enhancements.\\

\subsubsection{\underline{Coding information : }}
\begin{center}
\begin{figure}[htp]
\vspace{-0.25in} 
   \begin{minipage}[b]{1\linewidth}
    \centerline{\includegraphics[width=12cm]{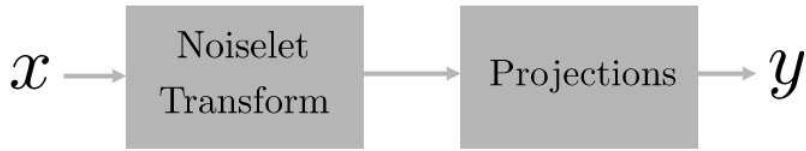}}
    \end{minipage}
\vspace{-0.5in} 
\caption{\textbf{The coding scheme}.}
 \label{fig:coding}
\end{figure}

\end{center}

The second way of applying compressed sensing for astronomical remote sensing is more conventional. As illustrated in Figure~\ref{fig:coding}, the coding stage mainly computes a few projections of the signal $x$. For the sake of economy, computing these projections should be computationally cheap. As stated in Section~\ref{sec:csgist}, good measurements vectors must be incoherent with the basis $\bf \Phi$ in which $x$ is assumed to be sparse. Fortunately, most astronomical data are sparsely represented in a wide range of wavelet bases. In that context, as emphasized by Cand\`es in \cite{Candes:2007lq}, noiselets (see \cite{Coifman01noiselets}) provide a near optimal measurement ensemble for astronomical data. 
The attractiveness of noiselets is twofold~:
\begin{itemize}
\item{Low computational cost :} on board compression can afford noiselet measurements as a fast transform (requiring $\mathcal{O}\left(t\right)$ flops) is available.
\item{Non-adaptive coding :} noiselets projections provide near-optimal measurements with most astronomical data that are sparsely represented in wavelet bases.
\end{itemize}
The coding process is non-adaptive~: the measurement ensemble $\bf \Theta$ may depend on the sparse representation $\bf \Phi$ but not directly on the data $x$. In this context, the measurement ensemble $\bf \Theta$ is efficient for a wide class of signals (sparse in the orthobasis $\bf \Phi$).

\subsection{Practical signal recovery}
\label{sec:astro_decoding}
In contrast to the simplicity of the coding step, the decoding step requires a more complex decompression scheme. As emphasized in Section~\ref{sec:sect_sensing}, the decoding step is equivalent to solving the inverse problem in Equation~\eqref{eq:l1_app}. Practical situations involving large scale problems require the use of a fast and accurate decoding algorithm. In this Section, we introduce a new fast algorithm for solving the recovery problem in Equation~\eqref{eq:l1_app}. We particularly focus on the flexibility of the decoding step. Indeed, in the compressed sensing framework, the decompression step can account for physical priors thus entailing higher performances.

\begin{center}
\begin{figure}[htb]
    \begin{minipage}[b]{1\linewidth}
    \centerline{\includegraphics[width=12cm]{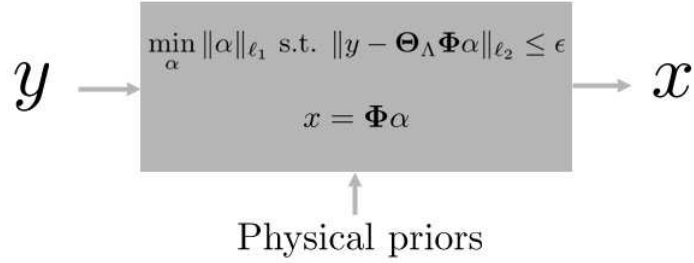}}
   \end{minipage}
\vspace{-0.1in} \caption{\textbf{The decoding scheme}.} \label{fig:decoding}
\end{figure}
\end{center}

\subsubsection{A practical and effective CS decoding algorithm}

The decoding or recovery step amounts to solving the following convex program~:
\begin{equation}
\label{eq:l1_app2}
\min_{\alpha} \|\alpha\|_{\ell_1} \mbox{ s.t. } \left\|y - {\bf \Theta}_{\Lambda} {\bf \Phi} \alpha \right\|_{\ell_2} \leq \epsilon
\end{equation}  
The measurement matrix is composed of a subset $\Lambda$ indexing $M=\mbox{Card}\left(\Lambda \right)$ row vectors of the orthonormal matrix $\bf \Theta$. Let define ${\bf I}_\Lambda$ as the diagonal matrix the entries of which are defined as follows~:
\begin{equation}
\forall i \in \{1,\cdots,t\}; \quad {\bf I}_\Lambda[i,i] =
\left\{
\begin{array}{cc}
1 & \mbox{ if } i \in \Lambda \\
0 & \mbox{ otherwise}
\end{array}
\right.
\end{equation}
where ${\bf I}_\Lambda[i,i]$ is the $i$-th diagonal element of ${\bf I}_\Lambda$. Let define the signal $y^\sharp$ of size $t$ as follows~:
\begin{equation}
 y^\sharp_\Lambda  =  y \mbox{ and } y^\sharp_{\Lambda^c}  =  0 \\
\end{equation}
where $\Lambda^c$ is complement of $\Lambda$ in $\{1,\cdots,t\}$.  The problem in Equation~\eqref{eq:l1_app2} is then recast as follows~:
\begin{equation}
\label{eq:l1_app3}
\min_{\alpha} \|\alpha\|_{\ell_1} \mbox{ s.t. } \left\|y^\sharp - {\bf I}_{\Lambda} {\bf \Theta} {\bf \Phi} \alpha \right\|_{\ell_2} \leq \epsilon
\end{equation}
With an appropriate bijective re-parametrization, there exists a constant $\gamma$ such that the problem in Equation~\eqref{eq:l1_app3} can be formulated as an augmented Lagrangian~:
\begin{equation}
\label{eq:lag}
\alpha = \Argmin{\alpha} \frac{1}{2}\left\|y^\sharp - {\bf I}_{\Lambda} {\bf \Theta} {\bf \Phi} \alpha \right\|_{\ell_2}^2 + \gamma \|\alpha\|_{\ell_1}
\end{equation}
A wide range of optimization techniques, often based on iterative thresholding, have been proposed to solve this problem ( \cite{rest:daube04,Fornasier:2007bh} to quote a few). Recently, a general framework \cite{CombettesWajs05} for solving such problems has been introduced based on proximal projections. In the light of the proximal forward-backward optimization techniques developed in \cite{CombettesWajs05}, solving the problem in Equation~\eqref{eq:lag} can be done by means of projected Landweber iterative algorithm. At iteration $(h)$, the coefficients $\alpha$ would be updated as follows~:
\begin{equation}
\label{eq:land1}
\alpha^{(h)} = \mathcal{S}_\gamma \left\{\alpha^{(h-1)} + {\bf R} \left(y^\sharp - {\bf I}_{\Lambda} {\bf \Theta} {\bf \Phi}\alpha^{(h-1)}  \right) \right\}
\end{equation}
where $\mathcal{S}_\gamma$ is the soft-thresholding operator with threshold $\gamma$. $\bf R$ is a relaxation descent-direction matrix such that the spectral 
radius of ${\bf I} - {\bf M} {\bf I}_{\Lambda} {\bf \Theta} {\bf \Phi}$ is bounded above by $1$. Choosing ${\bf R} =   {\bf \Phi}^T {\bf \Theta}^T {\bf I}_{\Lambda}$ entails appreciable simplifications~:
\begin{equation}
\label{eq:land3}
\alpha^{(h)} = \mathcal{S}_\gamma \left\{{\bf \Phi^T} {\bf \Theta^T}\left[ y^\sharp - {\bf I}_{\Lambda^c}{\bf \Theta} {\bf \Phi}\alpha^{(h-1)}  \right] \right\}
\end{equation}
Convergence conditions are given in \cite{CombettesWajs05}.

\paragraph{Choosing the regularization parameter $\gamma$}

The choice of the regularization parameter is crucial as it balances between the sparsity constraint and the how the solution fits the data. Classical approaches would advocate the use of cross-validation to estimate an optimal value of $\gamma$. Nevertheless, cross-validation is computationally expensive and thus not appropriate for large scale problems.\\
From a different point of view, solving the initial problem in Equation~\eqref{eq:l1_app2} can be done, under mild conditions, by homotopy continuation techniques (see \cite{Osborne2000,Efron04,donoho:l1greedy} and references therein). Such techniques iteratively selects coefficients $\alpha$ by managing active sets of coefficients. This kind of process has the flavor of iterative hard-thresholding with decreasing threshold $\gamma$. Inspired by such techniques, the threshold $\gamma$ is decreased at each iteration. It starts from $\gamma^{(0)} = \|{\bf \Phi^T} {\bf \Theta^T} y^\sharp \|_\infty$ and decreases towards $\gamma_{\min}$. The value of $\gamma_{\min}$ is $0$ in the noiseless case. When noise corrupts the data $y^\sharp$, $\gamma_{\min}$ may depend on the noise level. In Section~\ref{sec:results}, numerical results are given. In these experiments, noise contamination is assumed to be white Gaussian with zero mean and variance $\sigma_n^2$. In this case, the final threshold is chosen as $\gamma_{\min} = 3 \sigma_n$ which gives an upper bound for noise coefficients with overwhelming probability.\\
In practice, substituting the soft-thresholding operator in Equation~\eqref{eq:land3} by the hard thresholding operator provides better recovery performances. In the forthcoming experiments, we use hard-thresholding rather than soft-thresholding. 

\paragraph{The ProxIT algorithm}

The next panel introduces the ProxIT algorithm.

\begin{flushleft}
\vspace{0.025in}
\centering
\begin{tabular}{|c|} \hline
\begin{minipage}{1\linewidth}
\vspace{0.025in}
\footnotesize{\textsf{1. Set the number of iterations $I_{\max}$ and threshold $\gamma^{(0)} = \|{\bf \Phi^T} {\bf \Theta^T} y^\sharp \|_\infty$. $x^{(0)}$ is set to zero.}

\textsf{2. While  $\gamma^{(h)}$ is higher than a given lower bound $\gamma_{\min}$}

\hspace{0.25in} \textsf{$\bullet$ Compute the measurement projection of $x^{(h-1)}$:}

\hspace{0.5in} \textsf{$y^{(h)} = {\bf I}_{\Lambda}{\bf \Theta}x^{(h-1)}$.}

\hspace{0.25in} \textsf{$\bullet$ Estimate the current coefficients $\alpha^{(h)}$:}

\hspace{0.5in} \textsf{$\alpha^{(h)} = \mathcal{S}_{\gamma^{(h)}} \left\{{\bf \Phi^T} {\bf \Theta^T}\left[ y^\sharp - y^{(h)}  \right] \right\}$.}

\hspace{0.25in} \textsf{$\bullet$ Get the new estimate of $x$ by reconstructing from the selected coefficients ${\alpha}^{(h)}$ :}

\hspace{0.5in} \textsf{$x^{(h)} = {\bf \Theta} {\bf \Phi}\alpha^{(h)}$.}

\textsf{3. Decrease the threshold $\gamma^{(h)}$ following a given strategy.}}
\vspace{0.05in}
\end{minipage}
\\\hline
\end{tabular}
\vspace{0.05in}
\end{flushleft}

\paragraph{Remark}
Hereafter we enlighten some links between the ProxIT algorithm and previous work.\\
When, the measurement ensemble is the canonical basis of $\mathbb{R}^t$ (${\bf \Theta}={\bf I}$), the problem in Equation~\eqref{eq:l1_app3} can be equivalently rewritten as follows~:
\begin{equation}
\min_{\alpha} \|\alpha\|_{\ell_1} \mbox{ s.t. } \left\|y^\sharp - \mathcal{M}_{\Lambda} \odot x \right\|_{\ell_2} \leq \epsilon \mbox{ where } x = {\bf \Phi} \alpha
\end{equation}  
where $ \mathcal{M}_{\Lambda}$ is a binary mask of size $t$ such that~:
\begin{equation}
\forall i \in \{1,\cdots,t\}; \quad \mathcal{M}_\Lambda[i] =
\left\{
\begin{array}{cc}
1 & \mbox{ if } i \in \Lambda \\
0 & \mbox{ otherwise}
\end{array}
\right.
\end{equation}
This very special case of compressed sensing if equivalent to an interpolation known as \textit{inpainting} (filling holes in $x$). Interestingly, the ProxIT algorithm has then the flavor of the MCA inpainting algorithm introduced in \cite{fadili:icip05}. From that viewpoint, the ProxIT generalizes the former algorithm to a wider range of measurement ensembles.\\

\paragraph{Recovery results}

\begin{center}
\begin{figure}[htb]
    \begin{minipage}[b]{0.3\linewidth}
    \centerline{\includegraphics[width=5cm]{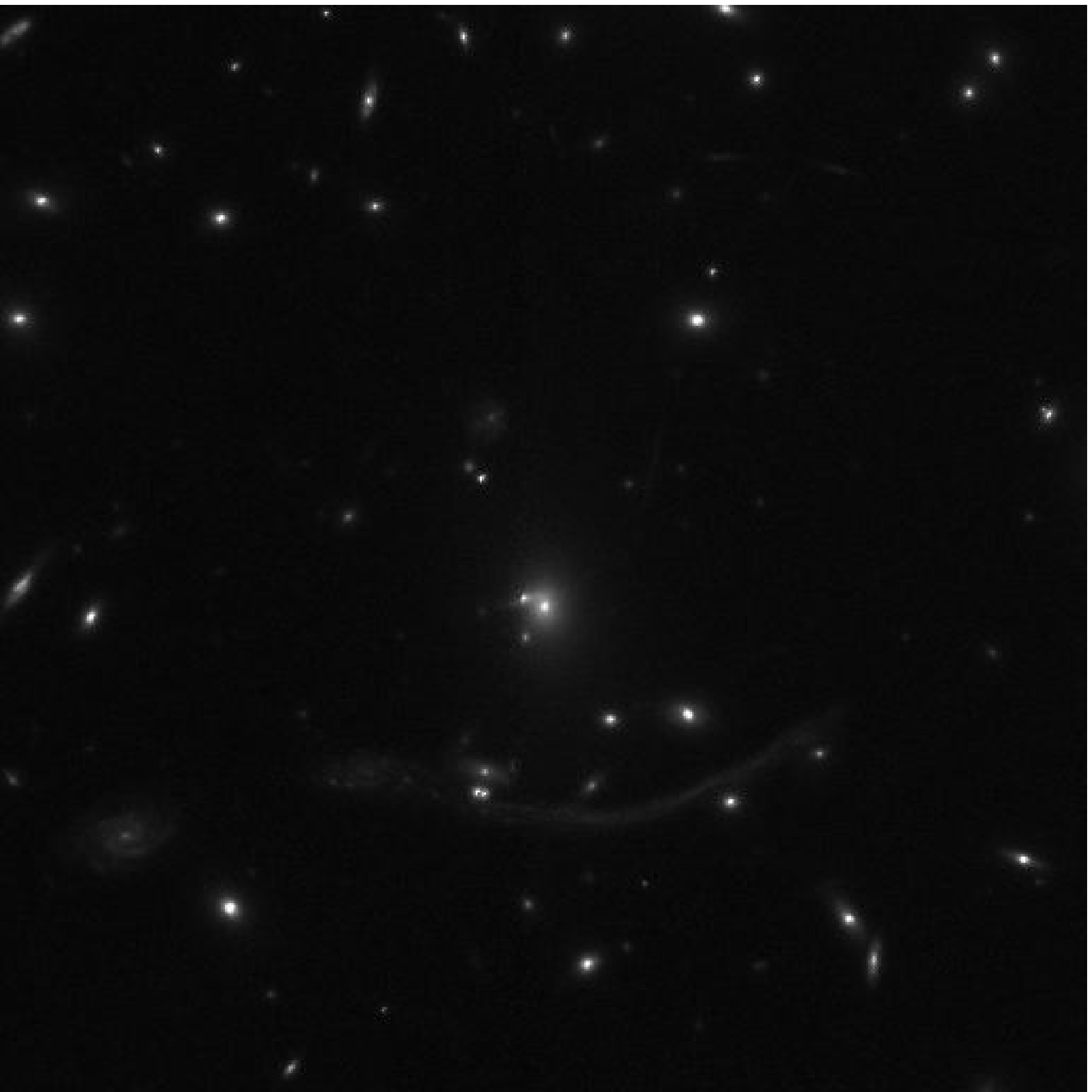}}
   \end{minipage}
   \hfill
   \begin{minipage}[b]{0.3\linewidth}
    \centerline{\includegraphics[width=5cm]{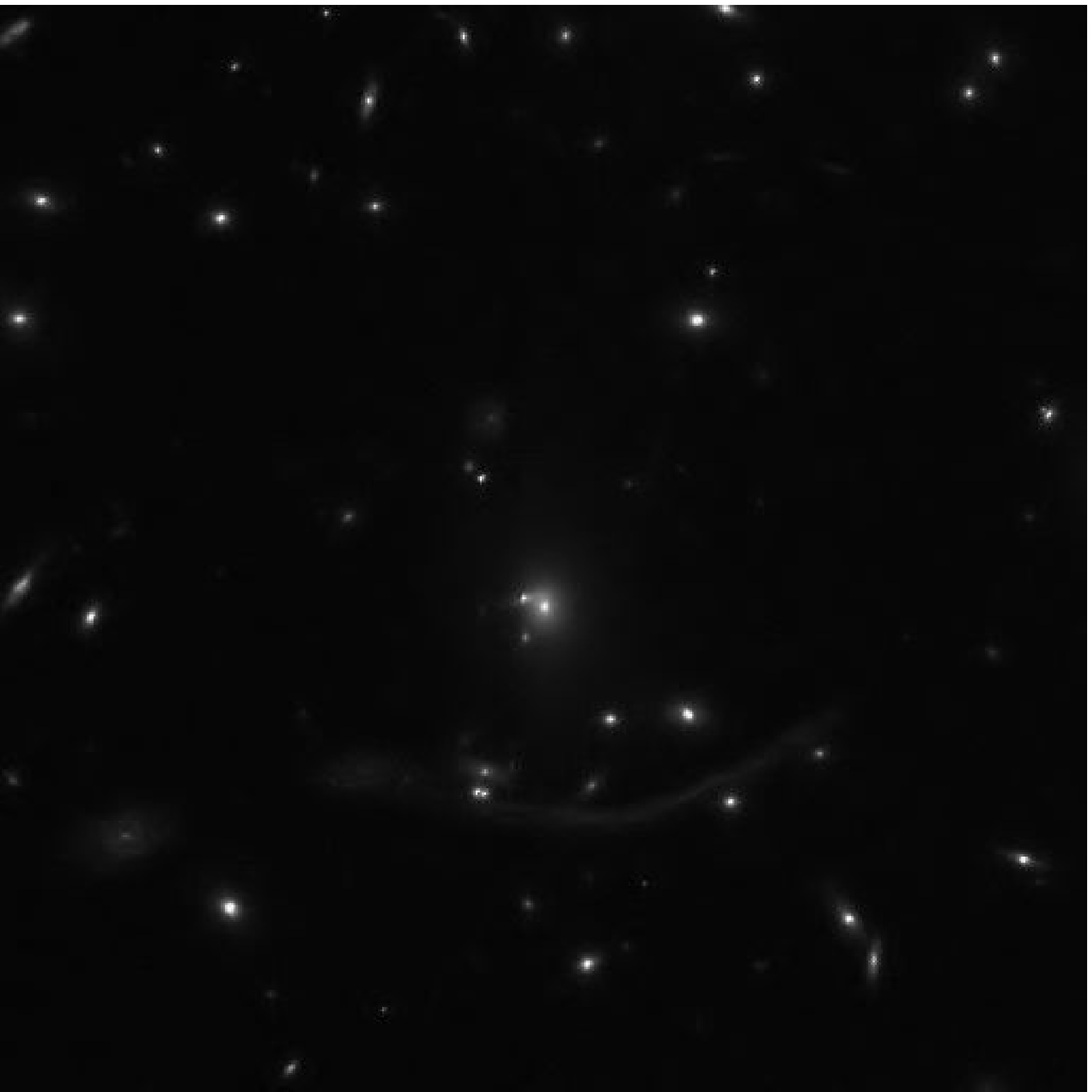}}
   \end{minipage}
   \hfill
   \begin{minipage}[b]{0.3\linewidth}
    \centerline{\includegraphics[width=5cm]{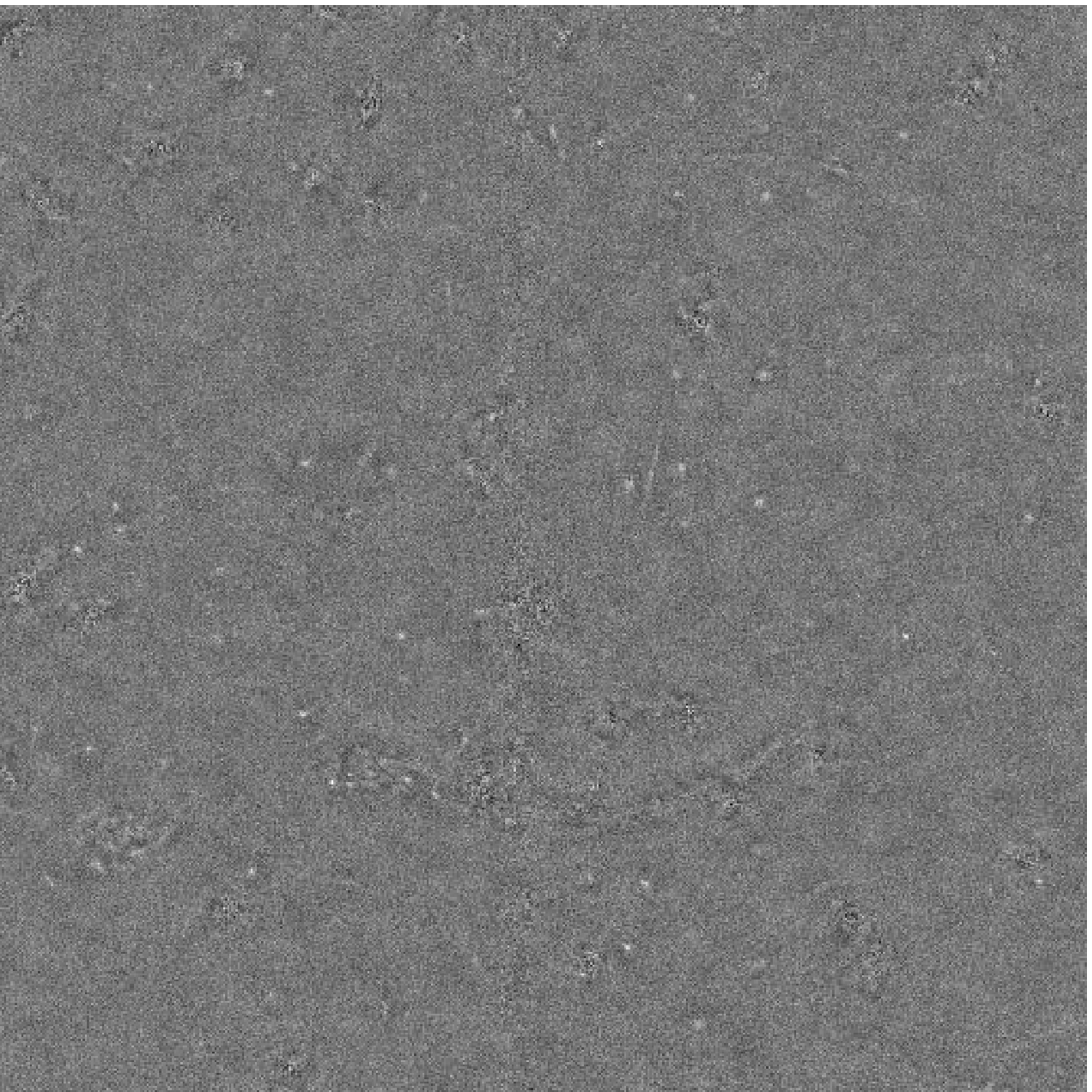}}
   \end{minipage}
\vspace{-0.1in} \caption{\textbf{Left :} Input image of size $512 \times 512$. \textbf{Middle :} Reconstruction from noiselet-based projections involving $20\%$ of the available projections. The ProxIT algorithm has been used with $P_{\max}=100$. \textbf{Right :} Difference between the original image and its CS-based reconstruction.} \label{fig:stcs1}
\end{figure}
\end{center}

\begin{center}
\begin{figure}[htb]
    \begin{minipage}[b]{0.3\linewidth}
    \centerline{\includegraphics[width=5cm]{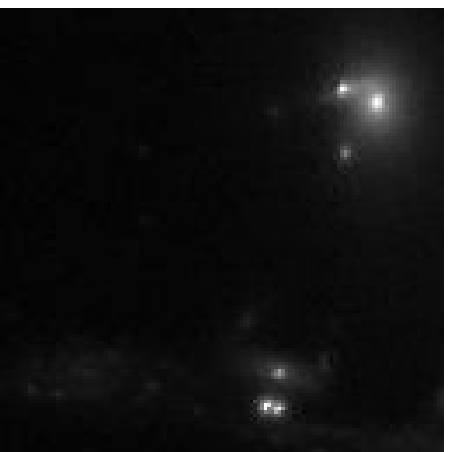}}
   \end{minipage}
   \hfill
   \begin{minipage}[b]{0.3\linewidth}
    \centerline{\includegraphics[width=5cm]{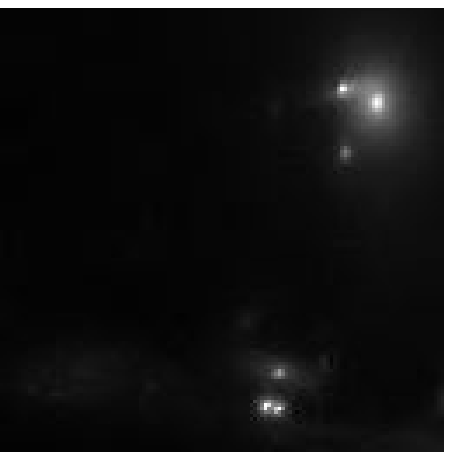}}
   \end{minipage}
   \hfill
   \begin{minipage}[b]{0.3\linewidth}
    \centerline{\includegraphics[width=5cm]{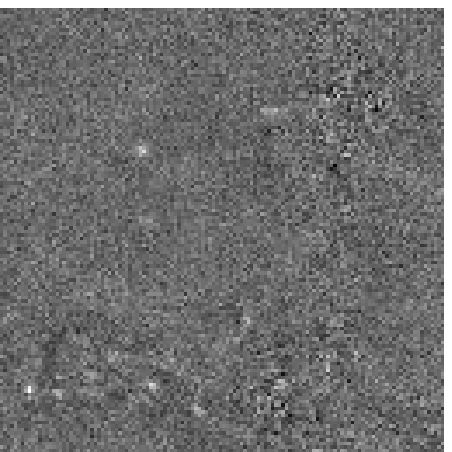}}
   \end{minipage}
\vspace{-0.1in} \caption{\textbf{Left :} Zoom of the input image of size $512 \times 512$. \textbf{Middle :} Zoom of the reconstruction from noiselet-based projections involving $20\%$ of the available projections ($\mbox{Card}(\Lambda)/t=0.2$). The ProxIT algorithm have been used with $P_{\max}=100$. \textbf{Right :} Zoom of the difference between the original image and its CS-based reconstruction.} \label{fig:stcs2}
\end{figure}
\end{center}

\begin{center}
\begin{figure}[htb]
    \begin{minipage}[b]{1\linewidth}
    \centerline{\includegraphics[width=12cm]{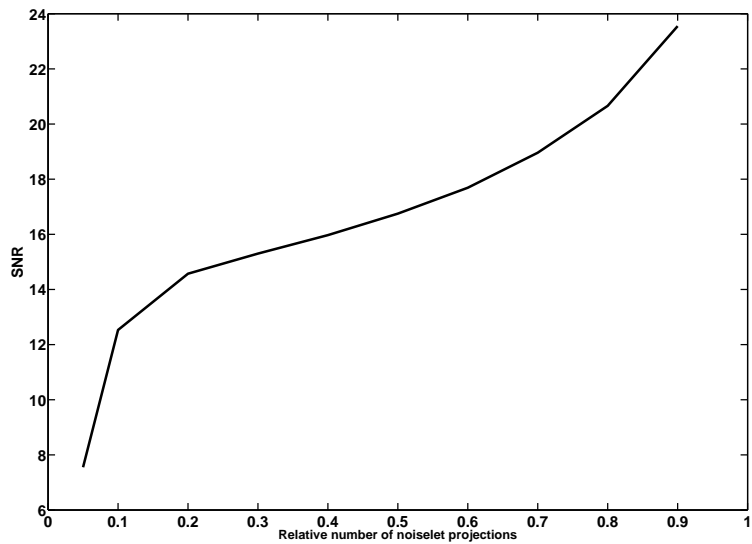}}
   \end{minipage}
\vspace{-0.1in} \caption{Recovery Signal-to-noise ratio when the relative number of noiselet projections varies.} \label{fig:recsnr}
\end{figure}
\end{center}

In this Section we provide several recovery results obtained using the ProxIT algorithm. In this experiment, the original data $x$ is a $512 \times 512$ HST\footnote{See \textit{http://~hubblesite.org/}.} image. Like most astronomical data, this signal is well (\textit{i.e.} sparsely) represented in a wavelet basis. Indeed, this kind of data mostly contains pointwise singularities (for instance stars or point sources) with smooth diffuse background. As stated earlier, choosing an effective measurement ensemble boils down to finding an orthobasis $\bf \Theta$ that is incoherent with the sparse representation $\bf \Phi$ (hereafter wavelets). Noiselets (see \cite{Coifman01noiselets}) are an orthogonal basis that is shown to be highly incoherent with a wide range of practical sparse representations (wavelets, Fourier to quote a few - see \cite{Candes:2007lq}). In the following experiment, the data $x$ are projected on a random subset of noiselet projections. More precisely, $y$ have been computed by randomly selecting coefficients of ${\bf \Theta}^Tx$. In the ProxIT algorithm, the sparse representation $\bf \Phi$ is an undecimated wavelet transform. The left picture of Figure~\ref{fig:stcs1} shows the original signal $x$. The picture in the middle features the signal $x$ recovered using the ProxIT algorithm from $0.2*t$ random noiselet projections. Pictures in Figure~\ref{fig:stcs2} depict the zoomed version of these images. Visually, the ProxIT algorithm performs well as it provides solutions close to the original data $x$. Both the pointwise structures and more diffuse features (such as the gravitational arc visible in Figure~\ref{fig:stcs2}) are effectively restored. 
The ProxIT algorithm has been performed on compressed signals with varying relative number of noiselet projections (compression rate) $\rho = \mbox{Card}\left(\Lambda\right)/t$. Figure~\ref{fig:recsnr} features the SNR of the recovery results when $\rho$ varies from $0.05$ to $0.9$. The ProxIT algorithm provides reasonable solutions for compression rate higher than $\rho = 0.1$. This experiment has been performed to enlighten the efficiency of the ProxIT algorithm for compressed sensing recovery issues. Performance analysis in the framework of the Herschel project  are presented in Section~\ref{sec:results}.

\subsubsection*{Comparison with other methods}
\begin{itemize}
\item{\underline{Linear programming :}} in the seminal paper \cite{wave:donoho98}, the authors proposed to solve the convex $\ell_1$-sparse decomposition problem in Equation~\eqref{eq:l1_app2} with linear programming methods such as interior point methods. Several techniques based on linear programming have been developed (see \cite{l1magic,l1ls} to name a few).Unfortunately, linear programming-based methods are computationally demanding and thus not well suited 
to large-scale problems such as ours.
\item{\underline{Greedy algorithms :}} the most popular greedy algorithm must be the Matching Pursuit and its orthogonal version OMP \cite{mallat93matching}. Conditions have been given under which MP and OMP are proved to solve the $\ell_1$ and $\ell_0$ sparse decomposition problems \cite{TG:cs,GIG,Sp:GribVand}. Greedy algorithms have also been proposed by the statistics community for solving variable selection problems (LARS/LASSO see \cite{Efron04,lasso:tib}). Homotopy-continuation algorithms have also been introduced to solve the sparse decomposition problem \cite{Osborne2000,MCW05,Plumb06}. Interestingly, a recent work by Donoho \cite{dt:fl1} sheds light on the links between greedy algorithms such as OMP, variable selection algorithms and homotopy. Such greedy algorithms however suffer from high computational cost. 
\item{\underline{Iterative thresholding :}} recently, iterative thresholding algorithms have been proposed to mitigate the greediness of the aforementioned \textit{stepwise} algorithms. Iterative thresholding has first been introduced for solving sparsity-based inverse problems (see \cite{starck:sta94_1,rest:daube04,Rest:Nowak,CombettesWajs05}). Some techniques based on iterative thresholding have been devised for CS (see \cite{WDarbon07,ist:fnw,donoho:stgomp} and references therein). The attractiveness of the proposed ProxIT algorithm is its simplicity~: i) it is a fast algorithm as computing $\bf \Theta$, $\bf \Phi$ (\textit{resp.} ${\bf \Theta}^T$, ${\bf \Phi}^T$) is performed by using implicit fast synthesis (\textit{resp.} analysis) transforms; ii) the ProxIT algorithm can easily account for further constraints such as positivity.\\ 
\end{itemize}

\subsubsection*{Accounting for physical priors}
In this section, we assume that the data $x$ have been compressed using compressed sensing. The ``observed" data $y$ are then made of $M$ incoherent projections~: $y = {\bf \Theta}_{\Lambda}x$. In the compressed sensing framework, the conventional decompression scheme would require solving the problem in Equation~\eqref{eq:l1_app2}. In real-world applications, further \textit{a priori} knowledge provides useful information to describe the data $x$. For instance, in astronomical applications, the data $x$ are often photon intensity. Positiveness is then a simple physical prior assumption to account for in the decoding step.\\
More generally, let assume that the useful data $x$ are observed through an ``observation" map $\mathcal{F}$; the compressed data $y$ are then recast as follows~:
\begin{equation}
y = {\bf \Theta}_{\Lambda} \mathcal{F}\left(x \right) + n
\end{equation}
where $n$ models projected instrumental noise or model imperfections. The ``observation" map $\mathcal{F}$ can model a wide range of physical or instrumental priors~: physical generating model, instrumental perturbations (convolution, instrumental detector response,$\cdots$etc.) to quote a few. In Section~\ref{sec:results}, the ``observation" map involves image shifts. In this context, accounting for such priors in the decoding step is desirable. The problem in Equation~\eqref{eq:l1_app2} is then rewritten as follows~:
\begin{equation}
\label{eq:l1_nlapp2}
\min_{\alpha} \|\alpha\|_{\ell_1} \mbox{ s.t. } \left\|y - {\bf \Theta}_{\Lambda} \mathcal{F}\left({\bf \Phi} \alpha\right) \right\|_{\ell_2} \leq \epsilon
\end{equation}
The ProxIT algorithm can be adapted to solve this problem. In case $\mathcal{F}$ is linear (\textit{i.e.} $\mathcal{F}\left(x \right) = Fx$ where $F$ is a $t \times t$ matrix - for instance, $F$ may model a convolution operator), extending the ProxIT algorithm to solve the problem in Equation~\eqref{eq:l1_nlapp2} is straightforward. In case $\mathcal{F}$ is non linear, the problem at hand gets far more difficult and will clearly depend on the expression of $\mathcal{F}$. Note that iterative thresholding-based techniques involving special instances of non-linear models have been studied in \cite{Teschke:2005eu}. In the next section, we will consider the case of bijective possibly non-linear maps $\mathcal{F}$.\\
To conclude this section, compressed sensing provides an attractive compression scheme~: i) the coding step is simple with a very low computational cost, ii) the decoding step is able to account for physical priors. Compressed sensing then fills the gap between data acquisition and data processing.

\subsection{Compressed sensing versus Standard compression techniques}
\label{sec:csomp}
CS-based compression have several advantages over standard compression techniques such as the celebrated JPEG\footnote{See \textit{http://www.jpeg.org/}.} compression standard.
\subsubsection{Computational complexity}
In case compressed sensing is used as a ``conventional" compression technique, the CS projections (noiselets in the forthcoming examples), require no further encoding in contrast to classical compression methods such as JPEG or JPEG2000. Furthermore, the only computational cost required by a CS-based compression is the computation of these projections. In case noiselets are used, their computational cost evolves as $\mathcal{O}(t)$ thus involving a low CPU load which is lower than the computational burden required by JPEG ($\mathcal{O}(t \log(t))$). It can be even much faster if these
projections are made with an optical system.

\subsubsection{Decoupling}
In contrast to classical compression techniques, there is a complete decoupling between the compression and the decompression in the CS framework.
Therefore the decompression step can be changed while keeping the same compressed data. This could is a very nice property. Indeed, we have seen
that the quality of the decompressed data is related to the sparsity of the data in a given basis $\bf \Phi$. If we discover in a few years 
a new dictionary which leads to a better sparsity of the data, then we can still improve the quality of the decompressed data.


\subsubsection{Data Fusion}
\begin{center}
\begin{figure}[htb]
    \begin{minipage}[b]{0.5\linewidth}
    \centerline{\includegraphics[width=5cm]{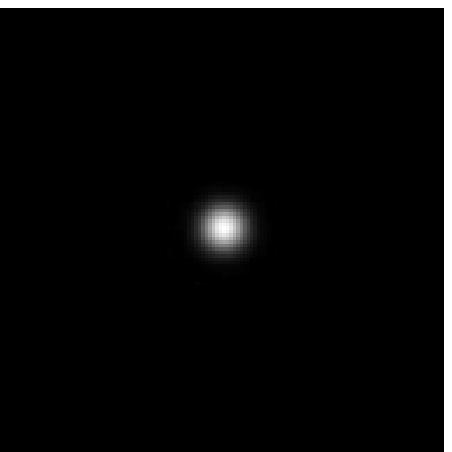}}
   \end{minipage}
   \hfill
   \begin{minipage}[b]{0.5\linewidth}
    \centerline{\includegraphics[width=5cm]{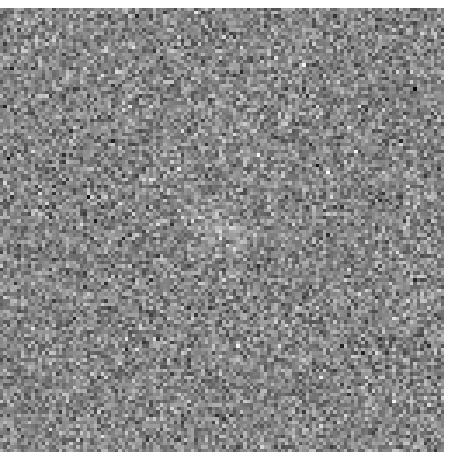}}
   \end{minipage}
   \vfill
   \vspace{0.1in}
   \begin{minipage}[b]{0.5\linewidth}
    \centerline{\includegraphics[width=5cm]{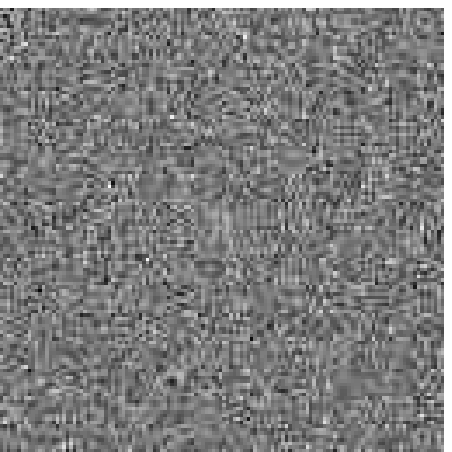}}
   \end{minipage}
   \hfill
   \begin{minipage}[b]{0.5\linewidth}
    \centerline{\includegraphics[width=5cm]{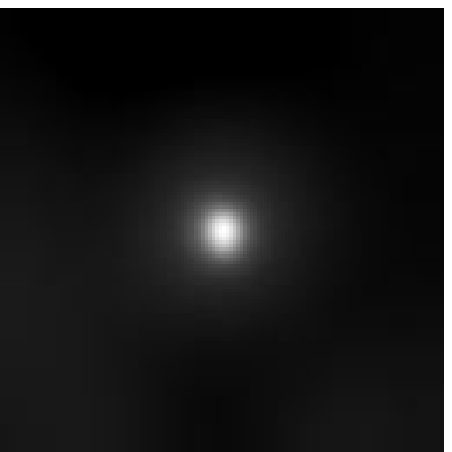}}
   \end{minipage}
\vspace{-0.1in} \caption{\textbf{Top - left :} Input image $x^\star$ of size $128 \times 128$. \textbf{Top - right :} First noisy input data $x_1$. White Gaussian noise is added with $SNR = -26$dB. \textbf{Bottom-left :} Estimate from the average of 10 images compressed by JPEG with a compression rate $\rho = 0.25$. \textbf{Bottom-right :} Estimate from $10$ pictures compressed by CS with a compression rate $\rho = 0.25$.} \label{fig:fps}
\end{figure}
\end{center}

In astronomy, remote sensing data involving specific scanning strategies (raster scans) often provide redundant information which cannot be accounted for by standard compression techniques. For instance, consider that the data are made of $10$ images $\left\{x_i \right\}_{i=1,\cdots,10}$ such that each image $x_i$ is the noisy version of the original datum $x^{\star}$~: $x_i = x^\star + n_i$ where $n_i$ is a white Gaussian noise with variance $\sigma_n^2 = 1$ and $\forall i\neq j; \quad \mathbb{E}\left\{n_i n_j\right\} = 0$. We assume that the original datum is a faint point source as depicted at the top on the left of Figure~\ref{fig:fps}. The SNR of each image $x_i$ is $-26$dB. The picture at the top on the right of Figure~\ref{fig:fps} depicts the first observed datum $x_1$. Each image $\left\{x_i \right\}_{i=1,\cdots,10}$ is compressed using JPEG and CS with a compression ratio $\rho = 0.25$. The picture at the bottom on the left of Figure~\ref{fig:fps} is the estimate of $x^\star$ which has been computed has the average of the $10$ compressed JPEG data. The picture at the bottom on the left in Figure~\ref{fig:fps} is the CS-based estimate of $x^\star$ which has been provided by using the ProxIT algorithm to solve the following decoding problem~:
\begin{equation}
\label{eq:fps}
\min_{\alpha^\star} \|\alpha^\star\|_{\ell_1} \mbox{ s.t. } \sum_{i=1}^{10} \left\|y_i -  {\bf \Theta}_{\Lambda} {\bf \Phi} \alpha^\star \right\|_{\ell_2} \leq \epsilon
\end{equation}  
where $x^\star = {\bf \Phi} \alpha^\star$ and $y_i = {\bf \Theta}_{\Lambda} x_i$. The measurement ensemble is made of noiselets. $\bf \Phi$ is an isotropic undecimated wavelet frame. Clearly, the JPEG compression leads to a catastrophic compression as the faint point source is not detectable after compression, while the CS-based compression technique is able to retrieve the faint point source as illustrated in Figure~\ref{fig:fps}. 

This huge difference for data fusion problems between both compression strategies 
is the consequence of a fundamental property of CS: {\bf the linearity of the compression}. In contrast to standard compression techniques (such as JPEG), the CS-based compression is linear. The data to transmit are indeed simple linear projections~: $y = {\bf \Theta}_{\Lambda} (x^\star + n)$ where $n$ models instrumental noise. Whatever the compression rate (\textit{i.e.} $\mbox{Card}\left(\Lambda\right)/t$), the incoherence between the measurement vectors ${\bf \Theta}_\Lambda$ and the data $x$ is likely to guarantee that $x^\star$ does not belong to the null space of ${\bf \Theta}_\Lambda$.  
As a consequence, the compressed data always contain a piece of information belonging to $x^\star$. 
 Standard compression methods (which are non-linear) do not verify this crucial property. For a faint source, a standard compression method will kill its noisy high frequencies and they will never 
be recovered whatever the number of times this source is observed. CS will increase the SNR of the source with growing number of observations.
Compressed sensing is flexible enough to take advantage (in the decompression step) of the redundancy of these kind of data to overcome the loss of SNR after compression.


\section{Experiment~: the Herschel project}
\label{sec:results}

Herschel is one of the cornerstone missions of the European Space Agency (ESA). This space telescope has been designed to observe in the far-infrared and sub-millimeter wavelength range. Its launch is scheduled for the fall of 2008. The shortest wavelength band, 57-210 $\mu$m, is covered by PACS (Photodetector Array Camera and Spectrometer) \cite{Poglitscha:2006oq}, which provides low to medium resolution spectroscopy and dual-band photometry. When PACS is used as a photometer, it will simultaneously image with its two bolometer arrays, a $64 \times 32$ and a $32 \times 16$ matrix, both read out at 40 Hz. The ESA is faced with a challenging problem~: conventional low-cost compression techniques cannot achieve a satisfactory compression rate. In this Section, we propose a new CS-based compression scheme for the Herschel/PACS data that yield an elegant and effective way to overcome the Herschel compression dilemma.

\subsection{The Herschel dilemma}

The Herschel space telescope is partially hampered by the narrowness of the transmission band compared to the large amount of data to be transferred. This handicap stems from the limitation of conventional compression techniques to provide adequate compression rate with low computational cost, given the high readout noise. More quantitatively, the data have to be compressed in real time by a factor of 16 with very low CPU power. The lossless compression (classically based on entropy coding) that is presently coded on board compresses the data by a factor of $2.5$. Up to now, the only acceptable solution (with respect to computational cost and quality) to overcome this need for a higher compression rate is the average of $i$ consecutive images, typically 6 \cite{Belbachira:2005}. For pointed observations this strategy is near-optimal as it increases the SNR by a factor of $\sqrt{i}$ without loss of spatial resolution. Moreover, computing the mean of $i$ images is clearly computationally very cheap.

Nevertheless, observing wide sky areas requires fast scanning strategies. In that case, the shift between consecutive images may reach approximately $\lambda = 1$ pixel while the FWHM (full width at half maximum) of the instrumental PSF (point spread function) is $\delta \simeq 3$ pixels. Averaging $6$ consecutive images yields an increase of the equivalent point spread function along the scanning direction thus leading to a loss of spatial resolution. This consequence can be catastrophic for some scientific programs. Furthermore, averaging is far less optimal for noise reduction as the useful part of the data is also spread when consecutive images are averaged. An effective compression scheme would have to balance between the following performance criteria~:
\begin{itemize}
\item{\bf Spatial resolution : } fast scan entails a low spatial resolution. An effective compression scheme would provide a lower resolution loss.
\item{\bf Sensitivity : } assuming that between consecutive non-shifted images instrumental noise is independent, averaging provides an optimal SNR. A lower noise ratio provides a higher signal detection ability. 
\end{itemize}
 
\subsection{Compressed sensing for the Herschel data}

The Herschel/PACS mission needs a compression rate equal to $6$. A first approach would amount to compress independently each image. As stated earlier, the more prior information is accounted for, the more effective the compression scheme is. Then, compressing $6$ consecutive images jointly would be more relevant. If we consider a stack of $6$ consecutive images $\{x_i\}_{i=0,\cdots,5}$, the simplest generative model is the following~:
\begin{equation}
\forall i \in \{0,\cdots,5\}; \quad x_i = \mathcal{T}_{\lambda_i}\left(x^\star\right) + n_i
\end{equation}
where $\mathcal{T}_{\lambda_i}$ is an operator that shifts the original image $x^\star$ with a shift $\lambda_i$. In practice, $x^{\star} = x_0$ and $\lambda_0 = 0$. The signal $n_i$ models instrumental noise or model imperfections. According to the compressed sensing framework, each signal is projected onto the subspace ranged by a subset of columns of $\bf \Theta$. Each compressed observation is then obtained as follows~:
\begin{equation}
\forall i \in \{0,\cdots,5\}; \quad y_i = {\bf \Theta}_{\Lambda_i} x_i
\end{equation} 
where the sets $\{\Lambda_i\}$ are such that~:
\begin{equation}
\sum_i {\bf  I}_{\Lambda_i} = {\bf I} \mbox{ and } \mbox{Card}\left( \Lambda_i \right) = C
\end{equation} 
The decoding step amounts to seeking the signal $x^\star$ as follows~:
\begin{equation}
\label{eq:l1_pacs}
\min_{x^\star} \|{\bf \Phi}^T \alpha\|_{\ell_1} \mbox{ s.t. } \sum_{i=1,\cdots,5} \left\|y_i - {\bf \Theta}_{\Lambda_i} \mathcal{T}_{\lambda_i}\left(x^\star\right) \right\|_{\ell_2}^2 \leq \epsilon^2 \mbox{ and } x^{\star} \ge 0
\end{equation} 
We propose solving this problem by using an adapted version of the ProxIT algorithm we introduced in Section~\ref{sec:stcs}. Furthermore, the content of astronomical data is often positive. Constraining the solution to be positive would help solving the recovery problem.  Assuming that the shifting operator $\mathcal{T}_{\lambda_i}$ is invertible\footnote{This assumption is true when shifting the image does note deteriorate the original signal.}, we substitute Equation~\eqref{eq:land3} by the following Equation\footnote{Note that if the operator $\mathcal{T}_{\lambda_i}$ were linear (\textit{i.e.} $\mathcal{T}_{\lambda_i}\left(x \right) = T_{\lambda_i} x$), then this update would be recast as follows~: ${x^\star}^{(h)} = \frac{1}{6} {\bf \Phi} \mathcal{S}_\gamma \left\{ {\bf \Phi}^T \sum_{i=1,\cdots,5}  T_{\lambda_i}^{-1} {\bf \Theta^T}\left[  y_i^\sharp - {\bf I}_{\Lambda_i^c}{\bf \Theta} T_{\lambda_i} {x{^\star}}^{(h-1)}  \right] \right\}$}~:
\begin{equation}
\label{eq:pacs3}
{x^\star}^{(h)} = \frac{1}{6} {\bf \Phi} \mathcal{S}_\gamma \left\{ {\bf \Phi}^T \sum_{i=1,\cdots,5}  \mathcal{T}_{- \lambda_i}\left( {\bf \Theta^T}\left[  y_i^\sharp - {\bf I}_{\Lambda_i^c}{\bf \Theta} \mathcal{T}_{\lambda_i}\left({x^\star}^{(h-1)}\right)  \right] \right)\right\}
\end{equation}
The positivity constraint is accounted for by projecting at each iteration the solution of the previous update equation on the cone generated by the vectors having positive entries~: ${x^\star}^{(h)} \leftarrow P_C\left({x^\star}^{(h)}\right)$ where the projector $P_C$ is defined as follows~:
\begin{equation}
\forall i=1,\cdots,t; \quad P_C\left(x\right)[i] =
\left\{
\begin{array}{cc}
x[i] & \mbox{ if } x[i] \ge 0 \\
0 & \mbox{ otherwise}
\end{array}
\right.
\end{equation}
where $P_C\left(x\right)[i]$ is the $i$-th entry of $P_C\left(x\right)$. 
In the next section, we illustrate the good performances of the proposed non-linear decoding scheme.

\begin{center}
\begin{figure}[htb]
    \begin{minipage}[b]{1\linewidth}
    \centerline{\includegraphics[width=14cm]{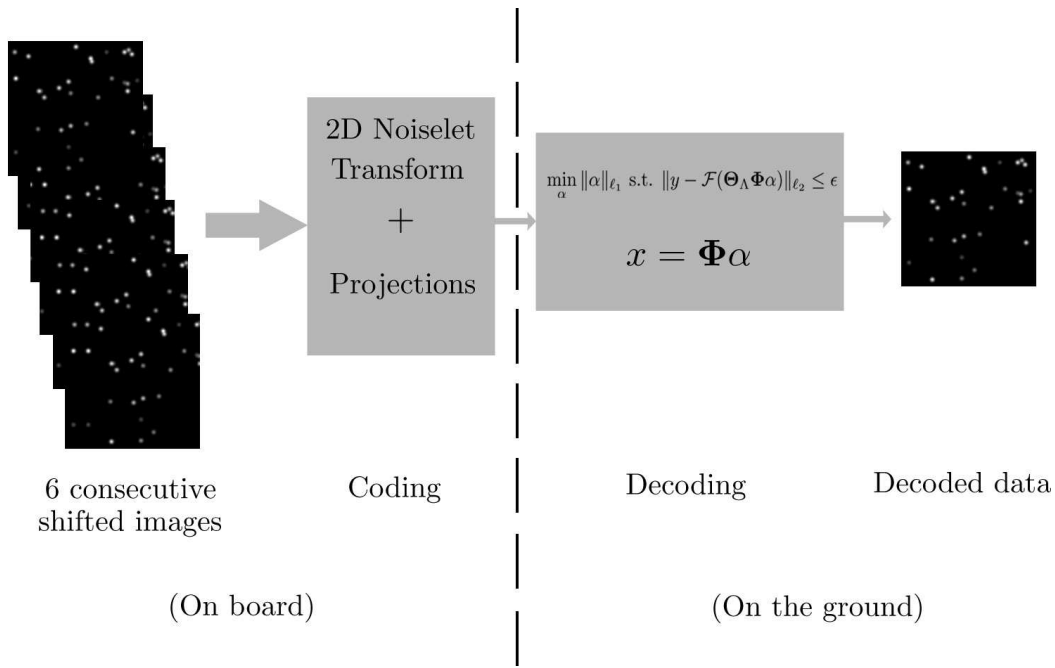}}
   \end{minipage}
\vspace{-0.1in} \caption{The proposed Herschel compression scheme.} \label{fig:hcscheme}
\end{figure}
\end{center}


\subsection*{Notations}

In the next experiments, the data will made of pointwise sources; it is worth defining some useful notations. Recall that we assume the telescope's PSF to have a FWHM equal to $\delta$. The shift between the original datum $x^\star$ and the $i$-th datum $x_i$ is $\lambda_i$. The intensity $f$ of the datum $x^\star$ is defined as its total flux~:
\begin{equation}
f = \sum_{j=1}^t x^\star[j]
\end{equation}
where $x[j]$ is the $j$-th entry. We also assume the $x^\star$ has positive entries.

\subsection{A toy-example}

In the following experiments, the datum $x^\star$ is a $128 \times 128$ image. The instrument is assumed to have a FWHM $\delta = 3$ pixels. For the sake of simplicity, each shift $\lambda_i = i$ pixels. White Gaussian noise is added to account for the instrumental noise.

\subsubsection{Detection performances}
\label{sec:detperf1}

\begin{center}
\begin{figure}[htb]
    \begin{minipage}[b]{0.5\linewidth}
    \centerline{\includegraphics[width=5cm]{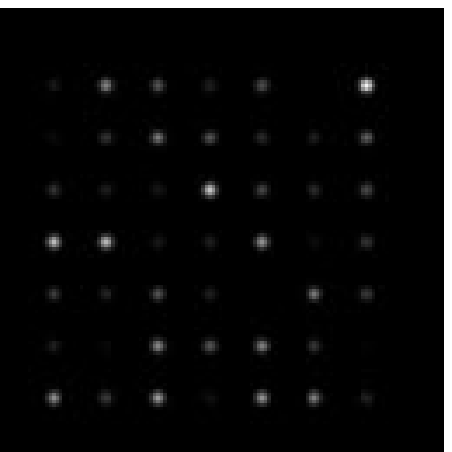}}
   \end{minipage}
   \hfill
   \begin{minipage}[b]{0.5\linewidth}
    \centerline{\includegraphics[width=5cm]{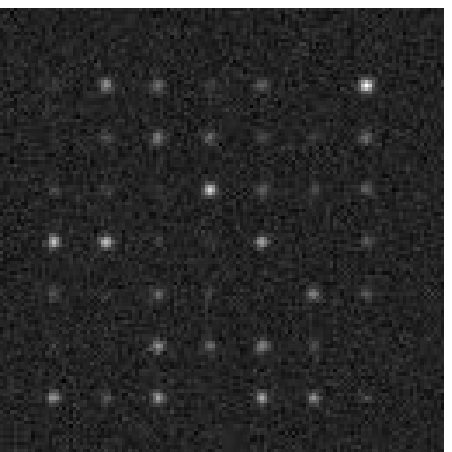}}
   \end{minipage}
   \vfill
   \vspace{0.1in}
   \begin{minipage}[b]{0.5\linewidth}
    \centerline{\includegraphics[width=5cm]{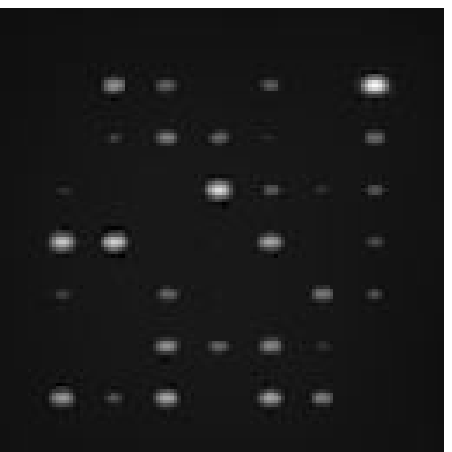}}
   \end{minipage}
   \hfill
   \begin{minipage}[b]{0.5\linewidth}
    \centerline{\includegraphics[width=5cm]{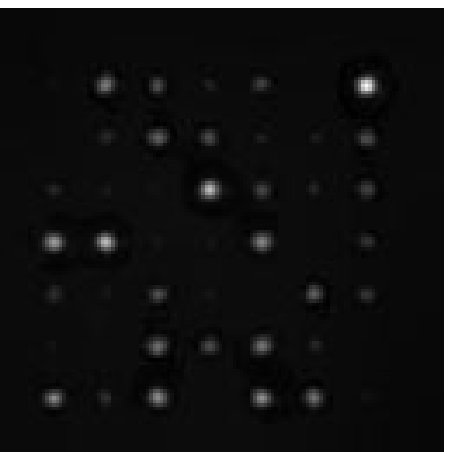}}
   \end{minipage}
\vspace{-0.1in} \caption{\textbf{Top left :} Original image of size $128 \times 128$ the total intensity of which is $f = 3500$. \textbf{Top right :} First input noisy map (out of $6$). White Gaussian with variance $\sigma^2_n = 1$ was added. \textbf{Bottom left :} Mean of the $6$ input images. \textbf{Bottom right :} Reconstruction from noiselet-based CS projections. The ProxIT algorithm has been used with $P_{\max}=100$.} \label{fig:detect1}
\end{figure}
\end{center}

\begin{center}
\begin{figure}[htb]
    \begin{minipage}[b]{1\linewidth}
    \centerline{\includegraphics[width=10cm]{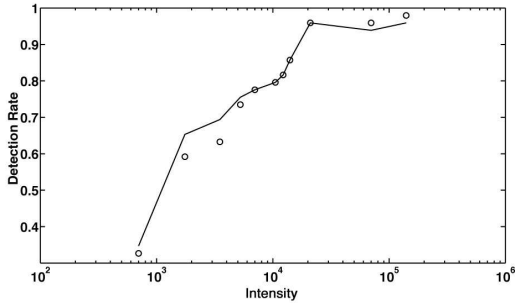}}
   \end{minipage}
\vspace{-0.1in} \caption{\textbf{Detection rate when the intensity of the input data varies : } \textit{Solid line} Resolution defined by the Rayleigh criterion of the CS-based reconstruction. \textit{$\bf \circ$} : Resolution of the solution provided by the mean of $6$ images.} \label{fig:detect2}
\end{figure}
\end{center}

In this experiment, the datum contains $49$ point sources that have been uniformly scattered. The amplitude of each point source is generated at random with a Gaussian distribution. The top-left picture of Figure~\ref{fig:detect1} shows the input data $x^\star$. The additive Gaussian noise has a fixed unit variance. The top-right panel of Figure~\ref{fig:detect1} features the data $x^\star$ contaminated with noise. Comparisons between the MO6 (``Mean of 6 images") and CS methods are made by evaluating for varying intensity value (from $700$ to $140000$; it is equivalent to a SNR varying from $-13.2$ to $33$dB) the rate of detected point sources. To avoid false detection, the same pre-processing step is performed~: i) ``\`a trous" bspline wavelet transform (see \cite{starck:book98}), ii) $5 \sigma_{M}$ hard-thresholding\footnote{Such  $5 \sigma_{M}$ is likely to avoid false detection as it defines a rather conservative threshold.} where $\sigma_{M}$ is the residual standard deviation estimated by a Median Absolute Deviation (MAD) at each wavelet scale, iii) reconstruction. The bottom-left panel of Figure~\ref{fig:detect1} features such filtered decoded image using the MO6 strategy. The bottom-right picture in Figure~\ref{fig:detect1}  shows the filtered ProxIT solution. In this experiment the total intensity of the point sources is set to $3500$. At first sight, both methods provide similar detection performances. As expected, the CS-based solution has a better spatial resolution.\\
Figure~\ref{fig:detect2} shows the detection rate (with no false detection) of each method for intensities varying from $f=700$ to $f=140000$. At high intensity (higher than $f=10^4$), both MO6 and CS provide rather similar detection performances. Interestingly, at low intensity, CS provides slightly better results. This unexpected phenomenon is partly due to the spread that results from the average of shifted images.\\
MO6 is theoretically (for low shifts) near-optimal for point source detection. In contrast, this experiment shows that CS can provide similar or better detection performances than MO6.
 
\subsubsection{Resolution}
\label{sec:resperf1}
\begin{center}
\begin{figure}[htb]
    \begin{minipage}[b]{0.5\linewidth}
    \centerline{\includegraphics[width=5cm]{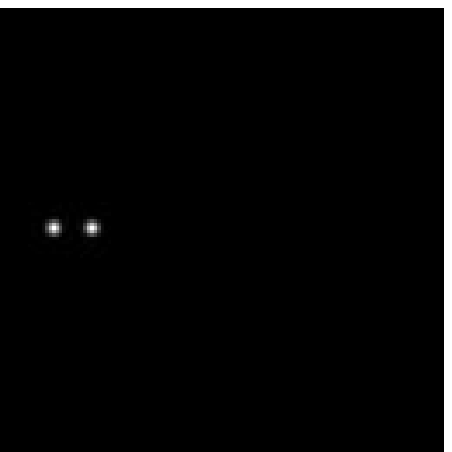}}
   \end{minipage}
   \hfill
   \begin{minipage}[b]{0.5\linewidth}
    \centerline{\includegraphics[width=5cm]{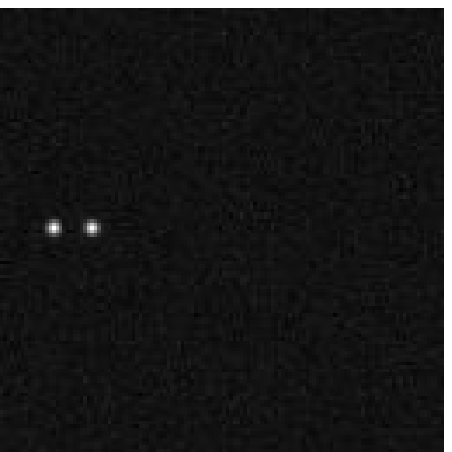}}
   \end{minipage}
   \vfill
   \vspace{0.1in}
   \begin{minipage}[b]{0.5\linewidth}
    \centerline{\includegraphics[width=5cm]{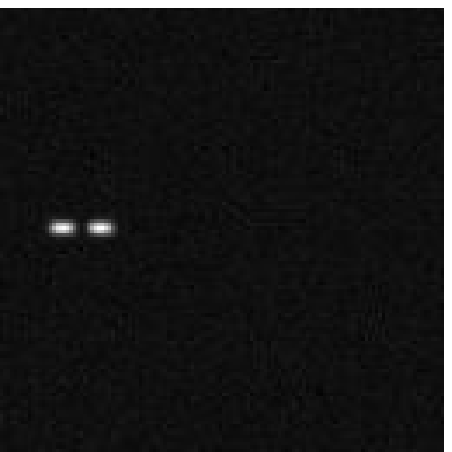}}
   \end{minipage}
   \hfill
   \begin{minipage}[b]{0.5\linewidth}
    \centerline{\includegraphics[width=5cm]{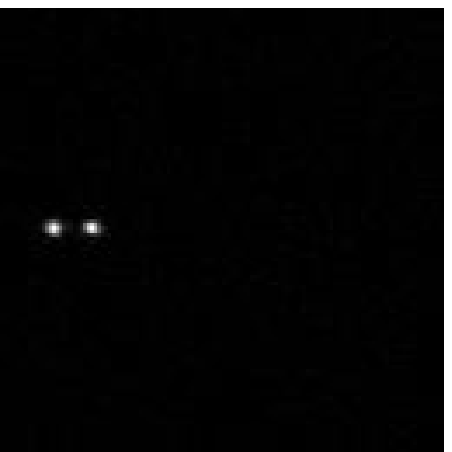}}
   \end{minipage}
\vspace{-0.1in} \caption{\textbf{Top left :} Original image of size $128 \times 128$ the total intensity of which is $f = 1000$. \textbf{Top right :} First input noisy map (out of $6$). White Gaussian with variance $\sigma^2_n = 1$ was added. \textbf{Bottom left :} Mean of the $6$ input images. \textbf{Bottom right :} Reconstruction from noiselet-based CS projections. The ProxIT algorithm has been used with $P_{\max}=100$.} \label{fig:res1}
\end{figure}
\end{center}

\begin{center}
\begin{figure}[htb]
    \begin{minipage}[b]{0.3\linewidth}
    \centerline{\includegraphics[width=5cm]{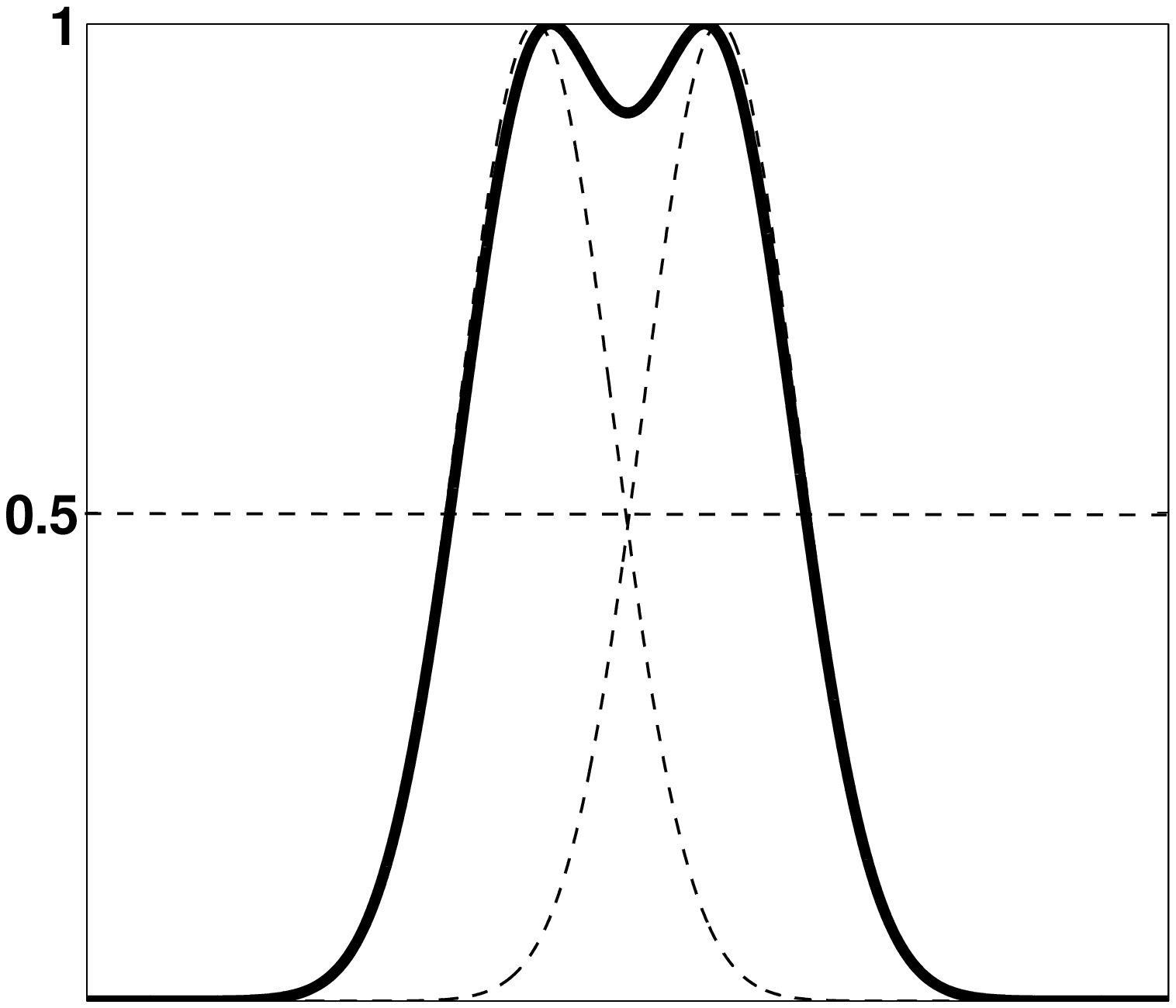}}
   \end{minipage}
   \hfill
   \begin{minipage}[b]{0.3\linewidth}
    \centerline{\includegraphics[width=5cm]{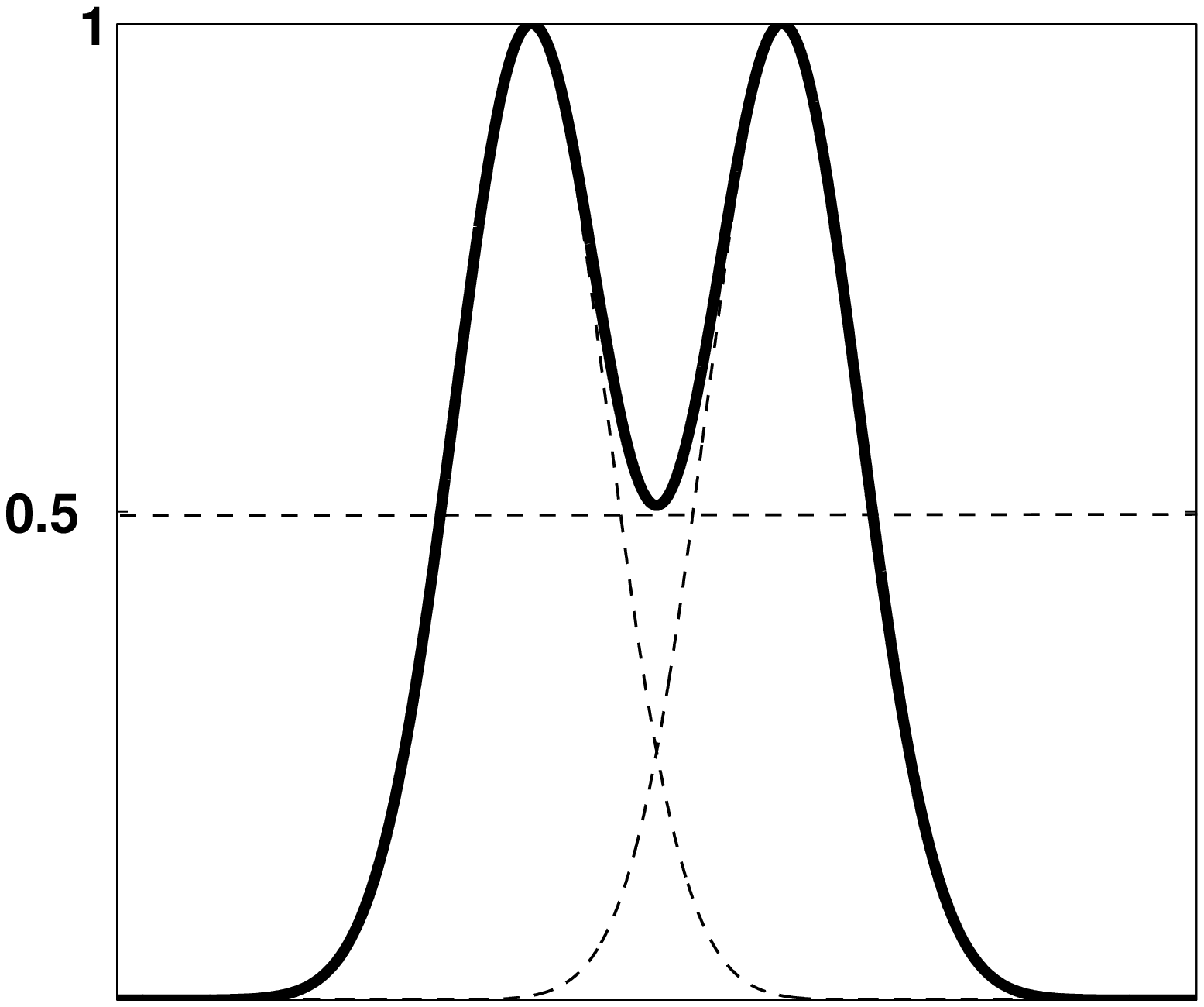}}
   \end{minipage}
   \hfill
   \vspace{0.1in}
   \begin{minipage}[b]{0.3\linewidth}
    \centerline{\includegraphics[width=5cm]{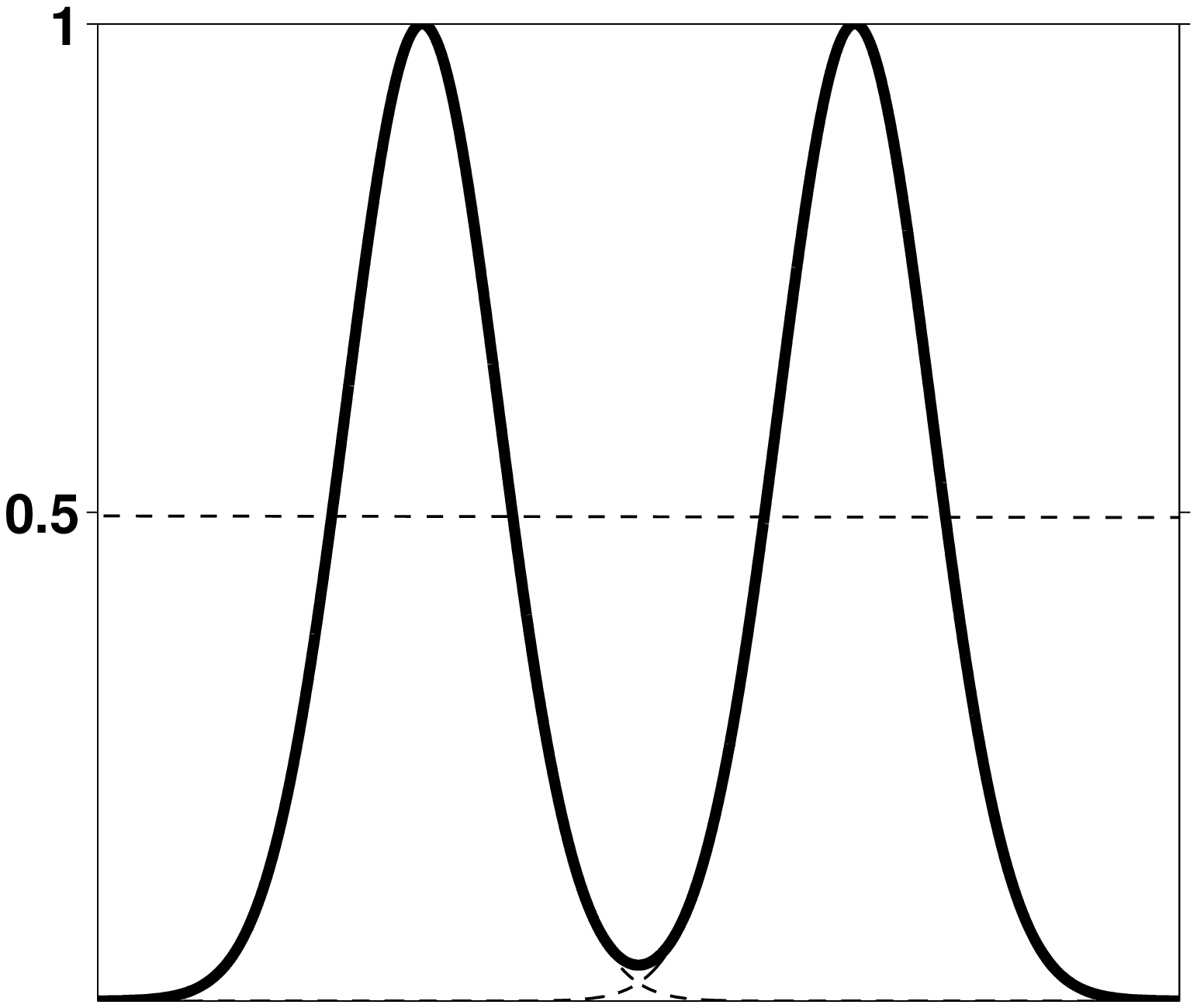}}
   \end{minipage}
\vspace{-0.1in} \caption{\textbf{The Rayleigh criterion - Left : } The point sources are not resolved. \textbf{Middle : } Resolution limit. \textbf{Right : } Fully resolved point sources.} \label{fig:rayl}
\end{figure}
\end{center}

Spatial resolution is a crucial instrumental feature. Averaging shifted images clearly deteriorates the final spatial resolution of Hershel/PACS. In this experiment, the original datum $x^\star$ is made of a couple of point sources. In the worst case, these point sources are aligned along the scan direction. The top-left picture of Figure~\ref{fig:res1} features the original signal $x^\star$. In the top-right panel of Figure~\ref{fig:res1}, the intensity of the point sources is set to $f=1000$ while the noise variance is $\sigma_{n}^2 =1$. The SNR of the data to compress is equal to $2.7$dB. The MO6 solution (\textit{resp.} the CS-based solution) is shown on the left (\textit{resp.} right) at the bottom of Figure~\ref{fig:res1}. As expected, the spatial resolution of the MO6 is clearly worse than the resolution of the input datum $x^\star$. Visually, the CS-based solution mitigate the resolution loss.\\
For different intensity of the datum $x^\star$ (from $100$ to $2000$), the spatial resolution is evaluated according to the Rayleigh criterion. The Rayleigh criterion is the generally accepted criterion for the minimum resolvable detail~: two point sources are resolved when the first minimum is lower than the amplitude at half maximum of a single point source as illustrated in Figure~\ref{fig:rayl}. For a fixed intensity $f$, the resolution limit is evaluated by seeking the minimal distance between the point sources for which the Rayleigh criterion is verified. For intensities varying from  $f = 100$ to $f=2000$, the resolution limit is reported in Table~1.  \\
The CS-based compression scheme provides a solution with better spatial resolution. At high intensity, the resolution gain (in comparison with MO6) is equal to a third of the instrumental FWHM ($1$ pixel). At low intensity, the resolution gain provided by the CS-based method slightly decreases.\\
This experiment shows that CS mitigates the resolution loss resulting from the joint compression of $6$ consecutive images. 

\begin{center}
\begin{table}[h!b!p!]
\begin{center}
\begin{tabular*}{1\textwidth}{@{\extracolsep{\fill}}|c||c|c|c|c|c|c|c|c|r|}
\hline
  \hline  
 SNR & $-17.3$  & $-9.35$ & $-3.3$  & $0.21$  & $2.7$  & $4.7$  & $6.2$  & $7.6$  & $8.7$  \\
 \hline
  Intensity & $100$  & $250$ & $500$  & $750$  & $1000$  & $1250$  & $1500$  & $1750$  & $2000$  \\
  \hline
   \hline
   MO6 & $2.7$  & $2.7$ & $2.7$  & $2.7$  & $2.7$  & $2.7$  & $2.7$  & $2.7$  & $2.7$  \\
  \hline
 CS & $2$  & $2$ & $1.7$  & $1.7$  & $1.7$  & $1.7$  & $1.7$  & $1.7$  & $1.7$  \\
   \hline
\end{tabular*} 
\caption{\textbf{Spatial resolution in pixels : } for varying datum flux, the resolution limit of each compression technique is reported. The CS-based compression entails a resolution gain equal to a $30\%$ of the spatial resolution provided by MO6.}
\end{center}
\end{table}\label{table1}
\end{center}


\subsection{Realistic data}

\subsubsection{The data}

Real Herschel/PACS data are more complex than those we simulated in the previous experiments. The original datum $x^\star$ is contaminated with a slowly varying ``flat field" component $c_f$. In a short sequence of $6$ consecutive images, the flat field component is almost fixed. In this context, the data $\{x_i\}_{i=0,\cdots,1}$ can then be modeled as follows~:
\begin{equation}
x_i = \mathcal{T}_{\lambda_i}\left(x^\star\right) + n_i + c_f
\end{equation}
Assuming that $c_f$ is known, the ProxIT algorithm can be updated by substituting Equation~\eqref{eq:pacs3} with the following~:
\begin{equation}
{x^\star}^{(h)} = \frac{1}{6} {\bf \Phi} \mathcal{S}_\gamma \left\{ {\bf \Phi}^T \sum_{i=1,\cdots,5}  \mathcal{T}_{- \lambda_i}\left( {\bf \Theta^T}\left[  y_i^\sharp - {\bf I}_{\Lambda_i^c}{\bf \Theta} \left(\mathcal{T}_{\lambda_i}\left({x^\star}^{(h-1)}\right) - c_f \right)  \right] \right)\right\}
\end{equation}
If $c_f$ is unknown, it can be estimated within the ProxIT algorithm. The next Section focuses on the resolution gain provided by the CS- based method in the scope of real Herschel/PACS data. The data have been designed by adding realistic pointwise sources to real calibration measurements performed in mid-2007. 

\subsubsection{Resolution}
\begin{center}
\begin{figure}[htb]
    \begin{minipage}[b]{0.5\linewidth}
    \centerline{\includegraphics[width=5cm]{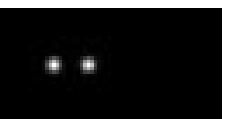}}
   \end{minipage}
   \hfill
   \begin{minipage}[b]{0.5\linewidth} 
    \centerline{\includegraphics[width=5cm]{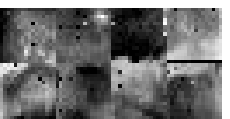}}
   \end{minipage}
   \vfill
   \vspace{0.1in}
   \begin{minipage}[b]{0.5\linewidth}
    \centerline{\includegraphics[width=5cm]{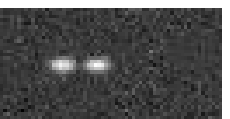}}
   \end{minipage}
   \hfill
   \begin{minipage}[b]{0.5\linewidth}
    \centerline{\includegraphics[width=5cm]{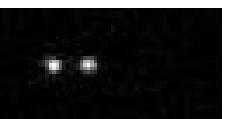}} 
   \end{minipage}
\vspace{-0.1in} \caption{\textbf{Top left :} Original image of size $32 \times 64$ with a total intensity of $f = 4500$. \textbf{Top right :} First input noisy map (out of $6$). The PACS data already contains approximately Gaussian noise. \textbf{Bottom left :} Mean of the $6$ input images. \textbf{Bottom right :} Reconstruction from noiselet-based CS projections. The ProxIT algorithm has been used with $P_{\max}=100$.} \label{fig:res_pacs}
\end{figure}
\end{center}

 Similarly to the experiments performed in Section~\ref{sec:resperf1}, we added a couple of point sources to Herschel/PACS data. The top-left picture of Figure~\ref{fig:res_pacs} features the original signal $x^\star$. In the top-right panel of Figure~\ref{fig:res_pacs}, the intensity of the point sources is set to $f=4500$. The ``flat field" component overwhelms the useful part of the data so that $x^\star$ has at best a level that is $30$ times lower than the ``flat field" component. The MO6 solution (\textit{resp.} the CS-based solution) is shown on the left (\textit{resp.} right) and at the bottom of Figure~\ref{fig:res_pacs} and all the results are presented in Table~2. Similarly to the previous fully simulated experiment, the CS-based algorithm provides better resolution performances. The resolution gain can reach $30\%$ of the FWHM of the instrument's PSF for a wide range of signal intensities. This experiment illustrates the reliability of the CS-based compression to deal with real-world data compression. 

\begin{center}
\begin{table}[h!b!p!]
\begin{center}
\begin{tabular*}{1\textwidth}{@{\extracolsep{\fill}}|c||c|c|c|c|c|c|c|c|r|}
\hline
  \hline  
 SNR & $-17.3$  & $-9.35$ & $-3.3$  & $0.21$  & $2.7$  & $4.7$  & $6.2$  & $7.6$  & $8.7$  \\
 \hline
  Intensity & $900$  & $2250$ & $4500$  & $6750$  & $9000$  & $11250$  & $13500$  & $15750$  & $18000$  \\
  \hline
   \hline
   MO6 & $3$  & $3$ & $3$  & $3$  & $3$  & $3$  & $3$  & $3$  & $3$  \\
  \hline
 CS & $2.33$  & $2.33$ & $2$  & $2$  & $2$  & $2$  & $2$  & $2$  & $2$  \\
   \hline
\end{tabular*} 
\caption{\textbf{Spatial resolution in pixels : } for varying datum flux, the resolution limit of each compression technique is reported. The CS-based compression entails a resolution gain equal to a $30\%$ of the spatial resolution provided by MO6.}
\end{center}
\end{table}\label{table2}
\end{center}

\section{Conclusion}
In this paper, we overview the potential applications of compressed sensing (CS) in astronomical imaging. The CS appeal in astronomy is twofold~: i) it provides a very easy and computationally cheap coding scheme for on-board astronomical remote sensing, ii)  the decoding stage is flexible enough to handle physical priors that lead to significant recovery enhancements. This paper introduces a new recovery algorithm to deal with the decoding problem. Based on iterative threshold, the ProxIT algorithm provides efficient approximate solutions to the decoding problem. Furthermore, the proposed algorithm is easy to handle as it requires setting only a few parameters. We show that the ProxIT algorithm is easily adapted to account for physical priors thus entailing better recovery results. We particularly point out the huge advantage of compressed sensing over standard compression techniques in the scope of multiple scanning observations (observing the same sky area several times). In this context, CS is able to provide astounding recovery results by taking advantage of the redundancy of the data.  We have shown that compressed sensing data fusion can lead to astounding improvements compared
to standard techniques. Preliminary numerical experiments illustrate the reliability of a CS-based compression scheme in the scope of astronomical remote sensing such as the Herschel space mission. We show that compressed sensing provides an elegant and effective compression technique that overcome the compression issue ESA is faced with.
In the next step we will focus on performing more realistic experiments in the scope of the Herschel space mission by adding more physical information.

\section{Acknowledgment}
The authors are very grateful to  E. Cand\`es for useful discussions and for having provided the noiselet code.

\bibliographystyle{IEEEtran}
\bibliography{AllBib,compress}

\begin{thebibliography}{10}
\providecommand{\url}[1]{#1}
\csname url@rmstyle\endcsname
\providecommand{\newblock}{\relax}
\providecommand{\bibinfo}[2]{#2}
\providecommand\BIBentrySTDinterwordspacing{\spaceskip=0pt\relax}
\providecommand\BIBentryALTinterwordstretchfactor{4}
\providecommand\BIBentryALTinterwordspacing{\spaceskip=\fontdimen2\font plus
\BIBentryALTinterwordstretchfactor\fontdimen3\font minus
  \fontdimen4\font\relax}
\providecommand\BIBforeignlanguage[2]{{%
\expandafter\ifx\csname l@#1\endcsname\relax
\typeout{** WARNING: IEEEtran.bst: No hyphenation pattern has been}%
\typeout{** loaded for the language `#1'. Using the pattern for}%
\typeout{** the default language instead.}%
\else
\language=\csname l@#1\endcsname
\fi
#2}}

\bibitem{compress:Veran}
J.~V\'eran and J.~Wright, ``Compression software for astronomical images,'' in
  \emph{Astronomical Data Analysis Software and Systems III}, D.~Worrall,
  C.~Biemesderfer, and J.~Barnes, Eds.\hskip 1em plus 0.5em minus 0.4em\relax
  Astronomical Society of the Pacific, 1994, p.~40.

\bibitem{compress:white92}
R.~White, M.~Postman, and M.~Lattanzi, ``Compression of the {G}uide {S}tar
  digitised {S}chmidt plates,'' in \emph{Digitized Optical Sky Surveys},
  H.~MacGillivray and E.~Thompson, Eds.\hskip 1em plus 0.5em minus 0.4em\relax
  Kluwer, 1992, pp. 167--175.

\bibitem{compress:press92}
W.~Press, ``Wavelet-based compression software for {FITS} images,'' in
  \emph{Astronomical Data Analysis Software and Systems I}, D.~Worrall,
  C.~Biemesderfer, and J.~Barnes, Eds.\hskip 1em plus 0.5em minus 0.4em\relax
  Astronomical Society of the Pacific, 1992, pp. 3--16.

\bibitem{starck:sta96_2}
J.-L. Starck, F.~Murtagh, B.~Pirenne, and M.~Albrecht, ``Astronomical image
  compression based on noise suppression,'' vol. 108, pp. 446--455, 1996.

\bibitem{compress:huang91_1}
L.~Huang and A.~Bijaoui, ``Astronomical image data compression by morphological
  skeleton transformation,'' \emph{Experimental Astronomy}, vol.~1, pp.
  311--327, 1991.

\bibitem{compress:dollet04}
C.~{Dollet}, A.~{Bijaoui}, and F.~{Mignard}, ``{All-sky imaging at high angular
  resolution: An overview using lossy compression},'' \emph{Astronomy and
  Astrophysics}, vol. 426, pp. 729--736, Nov. 2004.

\bibitem{CRT:cs}
E.~Cand\`es, J.~Romberg, and T.~Tao, ``Robust uncertainty principles: Exact
  signal reconstruction from highly incomplete frequency information,''
  \emph{IEEE Trans. on Information Theory}, vol.~52, no.~2, pp. 489--509, 2006.

\bibitem{CT:cs3}
E.~Cand\`es and T.~Tao, ``Near optimal signal recovery from random projections:
  Universal encoding strategies?'' \emph{IEEE Trans. on Information Theory},
  vol.~52, no.~12, pp. 5406--5425, 2006.

\bibitem{Donoho:cs}
D.Donoho, ``Compressed sensing,'' \emph{IEEE Trans. on Information Theory},
  vol.~52, no.~4, pp. 1289--1306, 2006.

\bibitem{Lustig07}
M.~Lustig, D.~Donoho, and J.~M. Pauly, ``Sparse mri: The application of
  compressed sensing for rapid mr imaging,'' \emph{Magnetic Resonance in
  Medicine}, vol.~58, no.~6, pp. 1182 -- 1195, 2007.

\bibitem{Sheikh07}
M.~Sheikh, O.~Milenkovic, and R.~Baraniuk, ``Designing compressive sensing dna
  microarrays,'' in \emph{EEE Workshop on Computational Advances in
  Multi-Sensor Adaptive Processing}, 2007.

\bibitem{Baradar07}
R.~Baraniuk and P.~Steeghs, ``Compressive radar imaging,'' in \emph{IEEE Radar
  Conference, Waltham, Massachusetts}, 2007.

\bibitem{LHerm07}
T.~Lin and F.~J. Herrmann, ``Compressed wavefield extrapolation,'' \emph{To
  appear in Geophysics}, 2007.

\bibitem{Candes:cs1}
E.~Cand\`es, ``Compressive sampling,'' \emph{International Congress of
  Mathematics, Madrid}, 2006.

\bibitem{Candes:2007lq}
W.~M. {Cand\`es}, E.~J., ````people hearing you without listening :" an
  introduction to compressive sampling,'' \emph{Preprint - available at
  \textit{http://www.dsp.ece.rice.edu/cs/}}, 2007.

\bibitem{cur:candes99_1}
E.~Cand\`es and D.~Donoho, ``Ridgelets: the key to high dimensional
  intermittency?'' \emph{Philosophical Transactions of the Royal Society of
  London A}, vol. 357, pp. 2495--2509, 1999.

\bibitem{curv:demanet}
E.~Cand\`es, L.~Demanet, D.~Donoho, and L.~Ying, ``Fast discrete curvelet
  transforms,'' \emph{SIAM Multiscale Model. Simul}, vol. 5/3, pp. 861--899,
  2006.

\bibitem{starck:sta01_3}
J.-L. Starck, E.~Cand\`es, and D.~Donoho, ``The curvelet transform for image
  denoising,'' vol.~11, no.~6, pp. 131--141, 2002.

\bibitem{cur:do05}
M.~N. Do and M.~Vetterli, ``The contourlet transform: an efficient directional
  multiresolution image representation,'' \emph{IEEE Transactions on Image
  Processing}, vol.~14, no.~12, pp. 2091--2106, 2005.

\bibitem{StarckCD02}
J.-L. Starck, E.~Cand{\`e}s, and D.~Donoho, ``The curvelet transform for image
  denoising,'' \emph{IEEE Transactions on Image Processing}, vol.~11, no.~6,
  pp. 670--684, 2002.

\bibitem{starck:sta05_4}
J.-L. Starck, M.~Elad, and D.~Donoho, ``Image decomposition via the combination
  of sparse representation and a variational approach,'' \emph{IEEE
  Transactions on Image Processing}, vol.~14, no.~10, pp. 1570--1582, 2005.

\bibitem{bobin:gmca_itip}
\BIBentryALTinterwordspacing
J.Bobin, J-L.Starck, J.Fadili, and Y.Moudden, ``Sparsity and morphological
  diversity in blind source separation,'' \emph{IEEE Transactions on Image
  Processing}, vol.~16, no.~11, pp. 2662 -- 2674, November 2007. [Online].
  Available: \url{http://perso.orange.fr/jbobin/pubs2.html}
\BIBentrySTDinterwordspacing

\bibitem{cur:vetterli01}
M.~Vetterli, ``Wavelets, approximation, and compression,'' \emph{IEEE Signal
  Processing Magazine}, vol.~18, no.~5, pp. 59--73, 2001.

\bibitem{Taubman:2001ij}
D.~{Taubman} and M.~{Marcellin}, \emph{JPEG2000: image compression
  fundamentals, standards and practice}.\hskip 1em plus 0.5em minus 0.4em\relax
  Kluwer, 2001.

\bibitem{Donoho-Bruckstein-Elad}
A.~Bruckstein, D.~Donoho, and M.~Elad, ``From sparse solutions of systems of
  equations to sparse modeling of signals and images,'' \emph{SIAM Review},
  2007, to appear.

\bibitem{DT:cs}
D.~Donoho and Y.~Tsaig, ``Extensions of compressed sensing,'' \emph{Signal
  Processing}, vol.~86, no.~3, pp. 5433--548, 2006.

\bibitem{CR:cs3}
E.~Cand\`es and J.~Romberg, ``Practical signal recovery from random
  projections,'' \emph{Preprint - available at
  \textit{http://www.dsp.ece.rice.edu/cs/}}, 2005.

\bibitem{miki:DonohoHuo}
D.~Donoho and X.~Huo, ``Uncertainty principles and ideal atomic
  decomposition,'' \emph{IEEE Trans. on Inf. Theory}, vol.~47, no.~7, pp.
  2845--2862, 2001.

\bibitem{CR:cs4}
E.~Cand\`es and J.~Romberg, ``Sparsity and incoherence in compressive
  sampling,'' \emph{Preprint - available at
  \textit{http://www.dsp.ece.rice.edu/cs/}}, 2006.

\bibitem{GIG}
J.~Tropp, ``Greedy is good : algorithmic results for sparse approximation,''
  \emph{IEEE Transactions on Information Theory}, vol.~50, no.~10, pp.
  2231--2242, 2004.

\bibitem{don:cs}
D.~L. {Donoho}, ``Compressed sensing,'' \emph{IEEE Trans. on Information
  Theory}, vol.~52, no.~4, pp. 1289--1306, 2006.

\bibitem{CR:cs2}
E.~Cand\`es and J.~Romberg, ``Quantitative robust uncertainty principles and
  optimally sparse decompositions,'' \emph{Foundations of Comput. Math},
  vol.~6, no.~2, pp. 227--254, 2006.

\bibitem{lasso:tib}
R.~Tibshirani, ``Regression shrinkage and selection via the lasso,'' \emph{J.
  R. Statist. Soc. B.}, vol.~58, no.~1, pp. 267--288, 1996.

\bibitem{wave:donoho98}
S.~Chen, D.~Donoho, and M.~Saunders, ``Atomic decomposition by basis pursuit,''
  \emph{SIAM Journal on Scientific Computing}, vol.~20, pp. 33--61, 1998.

\bibitem{CRT:cs2}
E.~Cand\`es, J.~Romberg, , and T.~Tao, ``Stable signal recovery from incomplete
  and inaccurate measurements,'' \emph{Communications on Pure and Applied
  Mathematics}, vol.~59, no.~8, pp. 1207--1223, 2006.

\bibitem{donoho:stgomp}
D.~Donoho, Y.~Tsaig, I.~Drori, and J.-L. Starck, ``Sparse solution of
  underdetermined linear equations by stagewise orthogonal matching pursuit,''
  \emph{IEEE Transactions On Information Theory}, 2006, submitted.

\bibitem{l1magic}
\BIBentryALTinterwordspacing
E.~J. {Cand\`es}, ``$\ell_1$-magic,'' Caltech, Tech. Rep., 2007. [Online].
  Available: \url{http://www.acm.caltech.edu/l1magic/}
\BIBentrySTDinterwordspacing

\bibitem{ist:fnw}
M.A.Figueiredo, R.~Nowak, and S.~Wright, ``Gradient projection for sparse
  reconstruction: Application to compressed sensing and other inverse
  problems,'' \emph{IEEE Journal of Selected Topics in Signal Processing - To
  appear}, 2007.

\bibitem{Skinner:2002qd}
G.~Skinner, ``Coded-mask imaging in gamma-ray astronomy - separating the real
  and imaginary parts of a complex subject,'' in \emph{Proceedings of the 22nd
  Moriond Astrophysics Meeting "The Gamma-Ray Universe"}, 2002.

\bibitem{TLW:cs}
D.~Takhar, J.~Laska, M.~Wakin, M.~Duarte, D.~Baron, S.~Sarvotham, K.~Kelly, and
  R.~Baraniuk, ``A new compressive imaging camera architecture using
  optical-domain compression,'' \emph{Proc. of Computational Imaging IV at SPIE
  Electronic Imaging, San Jose, California}, 2006.

\bibitem{Strong03}
A.~W. Strong, ``Maximum entropy imaging with integral/spi data,''
  \emph{Astronomy and Astrophysics}, vol. 411, no.~1, pp. L127--L129, 2003.

\bibitem{Coifman01noiselets}
R.~Coifman, F.~Geshwind, and Y.~Meyer, ``Noiselets,'' \emph{Appl. Comput.
  Harmon. Anal}, vol.~10, no.~1, pp. 27--44, 2001.

\bibitem{rest:daube04}
I.~Daubechies, M.~Defrise, and C.~D. Mol, ``An iterative thresholding algorithm
  for linear inverse problems with a sparsity constraint,'' \emph{Comm. Pure
  Appl. Math}, vol.~57, pp. 1413--1541, 2004.

\bibitem{Fornasier:2007bh}
\BIBentryALTinterwordspacing
M.~Fornasier and H.~Rauhut, ``Iterative thresholding algorithms,'' in
  \emph{Preprint}, 2007. [Online]. Available:
  \url{http://www.dsp.ece.rice.edu/cs/}
\BIBentrySTDinterwordspacing

\bibitem{CombettesWajs05}
P.~L. Combettes and V.~R. Wajs, ``Signal recovery by proximal forward-backward
  splitting,'' \emph{SIAM Journal on Multiscale Modeling and Simulation},
  vol.~4, no.~4, pp. 1168--1200, 2005.

\bibitem{Osborne2000}
M.~R. Osborne, B.~Presnell, and B.~A. Turlach, ``A new approach to variable
  selection in least squares problems,'' \emph{IMA Journal of Numerical
  Analysis}, vol.~20, no.~3, pp. 389--403, 2000.

\bibitem{Efron04}
B.~Efron, T.~Hastie, I.~Johnstone, and R.~Tibshirani, ``Least angle
  regression,'' \emph{Annals of Statistics}, vol.~32, no.~2, pp. 407--499,
  2004.

\bibitem{donoho:l1greedy}
D.~Donoho and Y.~Tsaig, ``Fast solution of $\ell_1$ minimization problems when
  the solution may be sparse,'' 2006, submitted.

\bibitem{fadili:icip05}
M.~J. Fadili and J.-L. Starck, ``Em algorithm for sparse representation - based
  image inpainting,'' \emph{IEEE International Conference on Image Processing
  ICIP'05}, vol.~2, pp. 61--63, 2005, genoa,Italia.

\bibitem{l1ls}
\BIBentryALTinterwordspacing
K.~Koh, S.-J. Kim, and S.~Boyd, ``Solver for l1-regularized least squares
  problems,'' Stanford University, Tech. Rep., 2007. [Online]. Available:
  \url{http://www.stanford.edu/$\sim$boyd/l1\_ls/}
\BIBentrySTDinterwordspacing

\bibitem{mallat93matching}
S.~Mallat and Z.~Zhang, ``Matching pursuits with time-frequency dictionaries,''
  \emph{IEEE Transactions on Signal Processing}, vol.~41, no.~12, pp.
  3397--3415, 1993.

\bibitem{TG:cs}
J.~Tropp and A.~Gilbert, ``Signal recovery from partial information via
  orthogonal matching pursuit,'' \emph{Preprint - available at
  \textit{http://www.dsp.ece.rice.edu/cs/}}, 2005.

\bibitem{Sp:GribVand}
R.Gribonval and P.Vandergheynst, ``On the exponential convergence of matching
  pursuits in quasi-incoherent dictionaries,'' \emph{IEEE Ttrans. Information
  Theory}, vol.~52, no.~1, pp. 255--261, 2006.

\bibitem{MCW05}
D.~M. Malioutov, M.~Cetin, and A.~S. Willsky, ``Homotopy continuation for
  sparse signal representation,'' in \emph{ICASSP '05}, vol.~5, 2005, pp.
  733--736.

\bibitem{Plumb06}
M.~Plumbley, ``Recovery of sparse representations by polytope faces pursuit,''
  in \emph{ICA06}, 2006, pp. 206--213.

\bibitem{dt:fl1}
D.~Donoho and Y.Tsaig, ``Fast solution of ell-1-norm minimization problems when
  the solution may be sparse.'' in \emph{Preprint available at
  http://www.dsp.ece.rice.edu/cs/}, 2006.

\bibitem{starck:sta94_1}
J.-L. Starck and F.~Murtagh, ``Image restoration with noise suppression using
  the wavelet transform,'' \emph{Astronomy and Astrophysics}, vol. 288, pp.
  343--348, 1994.

\bibitem{Rest:Nowak}
M.A.Figueiredo and R.D.Nowak, ``An em algorithm fo wavelet-based image
  restoration,'' \emph{IEEE Trans. On Image Processing}, vol.~12, no.~8, pp.
  906--916, August 2003.

\bibitem{WDarbon07}
\BIBentryALTinterwordspacing
W.~Yin, S.~Osher, D.~Goldfarm, and J.~Darbon, ``Bregman iterative algorithms
  for ell-1 minimization with applications to compressed sensing,'' Tech. Rep.,
  2007. [Online]. Available: \url{http://www.dsp.ece.rice.edu/cs/}
\BIBentrySTDinterwordspacing

\bibitem{Teschke:2005eu}
R.~Ramlau and G.~Teschke, ``A projection iteration for nonlinear operator
  equations with sparsity constraints,'' \emph{Numerische Mathematik}, vol.
  104, pp. 177--203, 2006.

\bibitem{Poglitscha:2006oq}
A.~Poglitscha, C.~Waelkensb, O.~Bauera, J.~Cepac, H.~Feuchtgrubera, T.~Henning,
  C.~van Hoofe, F.~Kerschbaumf, D.~Lemked, E.~Renotteg, L.~Rodriguez,
  P.~Saracenoi, and B.~Vandenbussche, ``The photodetector array camera and
  spectrometer (pacs) for the herschel space observatory,'' in \emph{SPIE},
  2006.

\bibitem{Belbachira:2005}
A.~N. Belbachir, H.~Bischof, R.~Ottensamer, F.~Kerschbaum, and C.~Reimers,
  ``On-board data processing to lower bandwidth requirements on an infrared
  astronomy satellite: Case of herschel-pacs camera,'' \emph{EURASIP Journal
  for Applied Signal Processing}, vol.~15, pp. 2585--2594, 2005.

\bibitem{starck:book98}
J.-L. Starck, F.~Murtagh, and A.~Bijaoui, \emph{Image Processing and Data
  Analysis: The Multiscale Approach}.\hskip 1em plus 0.5em minus 0.4em\relax
  Cambridge University Press, 2006.

\end{thebibliography}

\end{document}